# Semi-Supervised Approaches to Efficient Evaluation of Model Prediction Performance

Jessica Gronsbell & Tianxi Cai

## Abstract

In many modern machine learning applications, the outcome is expensive or time-consuming to collect while the predictor information is easy to obtain. Semi-supervised learning (SSL) aims at utilizing large amounts of 'unlabeled' data along with small amounts of 'labeled' data to improve the efficiency of a classical supervised approach. Though numerous SSL classification and prediction procedures have been proposed in recent years, no methods currently exist to evaluate the prediction performance of a working regression model. In the context of developing phenotyping algorithms derived from electronic medical records (EMR), we present an efficient two-step estimation procedure for evaluating a binary classifier based on various prediction performance measures in the semi-supervised (SS) setting. In step I, the labeled data is used to obtain a non-parametrically calibrated estimate of the conditional risk function. In step II, SS estimates of the prediction accuracy parameters are constructed based on the estimated conditional risk function and the unlabeled data. We demonstrate that under mild regularity conditions the proposed estimators are consistent and asymptotically normal. Importantly, the asymptotic variance of the SS estimators is always smaller than that of the supervised counterparts under correct model specification. We also correct for potential overfitting bias in the SS estimators in finite sample with cross-validation and develop a perturbation resampling procedure to approximate their distributions. Our proposals are evaluated through extensive simulation studies and illustrated with two real EMR studies aiming to develop phenotyping algorithms for rheumatoid arthritis and multiple sclerosis.

# 1. Introduction

Semi-supervised learning (SSL) has attracted much attention in the machine learning community in recent years. The typical semi-supervised (SS) set-up is characterized by two sources of data: (i) a small or moderate sized 'labeled' dataset $\mathcal{L}$ containing information on the outcome $y$ and the predictors $\mathbf{x}$ and (ii) a large 'unlabeled' dataset $\mathcal{U}$ containing information only on $\mathbf{x}$. This setting is directly relevant to many 'big data' applications where $y$ is difficult to obtain and $\mathbf{x}$ is readily available. For example, in the analysis of massive electronically recorded databases such as electronic medical records (EMR), it is easy to automatically extract $\mathbf{x}$, but time-consuming to manually label each observation with gold standard outcome information. SSL has thus proved applicable to a wide variety of similar practical problems including webpage classification (Liu et al., 2006; Wang and Chen, 2011), natural language parsing (Collobert and Weston, 2008; Søgaard, 2013), and object recognition (Rosenberg et al., 2005; Cheng et al., 2015).

As the name suggests, SSL differs from traditional supervised learning by making use of both $\mathcal{L}$ and $\mathcal{U}$ in the learning task. The primary interest of SSL is to determine if and when information contained in $\mathcal{U}$ can improve estimation precision relative to a supervised approach that ignores the unlabeled examples. Generally speaking, such improvement relies on a relationship between the parameter of interest and the distribution of $\mathbf{x}$ as $\mathcal{U}$ essentially characterizes the covariate distribution due to its size (Seeger, 2000; Zhang and Oles, 2000). As a motivating example, and to provide some intuition for the potential value of $\mathcal{U}$, consider the estimation of the population mean $\mu$ in the SS setting with a univariate predictor $x$. We have available the labeled dataset $\mathcal{L} = \{(y_i, x_i) : i = 1, ..., n\}$ and an independent unlabeled dataset $\mathcal{U} = \{x_i : i = n+1, ..., n+N\}$ with $N >> n$. Clearly, the supervised estimator of $\mu$ is $\overline{Y} = n^{-1}\sum_{i=1}^{n} y_i$. However, $\mu$ inherently depends on the marginal distribution of $x$ based on the fact that

$$\mu = E(Y) = E\{E(Y \mid X)\} = \int m(x)dF(x)$$

where $m(x) = E(Y \mid X = x)$ and $F(x) = P(X \leq x)$. Observing $\mathcal{U}$ gives additional information about $F(x)$ and in turn potential to improve the efficiency of $\overline{Y}$. A SS estimator of $\mu$ can correspondingly be constructed in two steps. First, $m(x)$ is estimated non-parametrically via kernel smoothing as

$$\widetilde{m}(x) = \frac{\sum_{i=1}^{n} K_h(x_i - x) y_i}{\sum_{i=1}^{n} K_h(x_i - x)}$$

where $K_h(u) = h^{-1}K(u/h)$, $K(\cdot)$ is a symmetric smooth density function, and $h$ is the bandwidth. Next, $\mu$ is estimated with

$$\widehat{\mu} = \int \widetilde{m}(x) d\widehat{F}(x)$$

where $\widehat{F}(x) = N^{-1} \sum_{i=n+1}^{n+N} I(x_i \leq x)$. Using arguments based on properties of the kernel estimator under standard smoothness assumptions and under-smoothing, one may easily show that (Bickel and Rosenblatt, 1973)

$$n^{\frac{1}{2}}(\widehat{\mu} - \mu) = n^{-\frac{1}{2}} \sum_{i=1}^{n} \{y_i - m(x_i)\} + o_p(1) \quad \text{while} \quad n^{\frac{1}{2}}(\overline{Y} - \mu) = n^{-\frac{1}{2}} \sum_{i=1}^{n} (y_i - \mu) + o_p(1).$$

The SS estimator therefore has asymptotic variance always smaller than that of $\overline{Y}$ provided $x$ is related to $y$.

Though the estimation of $\mu$ is simple example, it illustrates how knowledge of the marginal distribution of $\mathbf{x}$ can work to improve the efficiency of a supervised approach without the complexities that arise in more complicated settings. Within current literature, the SSL problem has been studied almost entirely in the context of estimating regression parameters or prediction rules (Chapelle et al., 2006; Zhu, 2006). Initially, it was assumed that "unclassified observations should certainly not be discarded" (O'neill, 1978). However, it has since been noted that unlabeled data can actually degrade estimation accuracy under model misspecification (Cozman et al., 2002, 2003; Grandvalet and Bengio, 2004; Singh et al., 2009). SS methods are thus motivated by 'safely' using unlabeled data to produce estimators that are always at least as efficient as supervised procedures regardless of model specification. Indeed, a wide range of approaches for classification and regression have been proposed including generative modeling, transductive methods, manifold regularization and graph regularization methods, all of which rely on implicit or explicit assumptions relating $p(\mathbf{x})$ to $p(y|\mathbf{x})$ to guarantee improvement over supervised learning (Baluja, 1998; Jaakkola et al., 1999; Belkin and Niyogi, 2004; Lafferty and Wasserman, 2006; Nigam et al., 2006; Belkin et al., 2006; Niyogi, 2008; Wasserman and Lafferty, 2008).

Despite this rich literature, no SSL methods currently exist to improve the estimation of prediction performance parameters. Recently, Claesen et al. (2015) proposed a procedure to estimate the bounds of the receiver operating characteristic (ROC) curve under the positive unlabeled setting in which $\mathcal{L}$ consists of examples from a single class. Their objective is therefore substantially different from improving the efficiency in estimating the accuracy measures in the standard SS setting where the labeled data consists of random samples from both classes, which is the goal of this paper. Such methods for accurately assessing the prediction performance of a working model have important implications in practice, particularly for EMR phenotyping. Recently, the Informatics for Integrating Biology and the Bedside (i2b2) Center, an NIH-funded National Center for Biomedical Computing, has developed several EMR phenotyping algorithms using supervised methods for identifying cases with rheumatoid arthritis (RA), multiple sclerosis (MS), Crohn's disease, and ulcerative colitis (Liao et al., 2010, 2012; Xia et al., 2013; Ananthakrishnan et al., 2013). A more precise method to evaluate these algorithms is especially valuable as rates of disease misclassification can have a profound impact on the power of clinical and genetic studies that require accurate disease definitions (Liao et al., 2010; Sinnott et al., 2014).

In this article, we address this gap in methodology through the development of an efficient SS procedure for estimating the accuracy of an estimated risk prediction rule for classifying a binary $y$. Specifically, we propose to use $\mathcal{L}$ and $\mathcal{U}$ together to efficiently estimate various accuracy parameters of the estimated prediction model, including the ROC curve, through a two-step procedure. In step I, $\mathcal{L}$ is used to obtain a non-parametrically calibrated estimate of the conditional risk function. In step II, SS estimators of the accuracy parameters are constructed by projecting the conditional risk function to the data in $\mathcal{U}$. This procedure bears similar intuition to the population mean example as the proposed estimators are functionals of the partial mean. Moreover, unlike previous work in SSL, our major contribution is the extension of the SS paradigm to the estimation of model evaluation metrics.

The remainder of this paper is organized as follows. In section 2, we formally introduce the SS set-up and the prediction performance measures of interest. In section 3, we formulate our estimators and detail the $\mathcal{K}$-fold cross-validation method used to correct for finite sample bias. We also present the perturbation resampling procedure for making inference and provide asymptotic properties for the SS estimators that confirm they are asymptotically more efficient than their supervised counterparts. In section 4, we demonstrate the validity of our proposals in finite sample with an extensive simulation study. We then illustrate the practical utility of our method in two EMR studies aiming to develop phenotyping algorithms for RA and MS in section 5. We conclude with additional discussions in section 6

# 2. Problem Set-Up

## 2.1. Data Representation in the SS Setting

With the development of phenotyping algorithms derived from EMRs as our motivating example, we let $y$ denote the binary phenotype of interest throughout. We denote by $\mathbf{x} = (1, x_1, \ldots, x_p)^\intercal$ the predictor vector for some fixed $p$. We also let $(y_0, \mathbf{x}_0{}^\intercal)^\intercal$ be the data vector for a future observation drawn independently from $p(y, \mathbf{x})$. The data for analysis in the SS setting is $\mathcal{D} = \mathcal{L} \cup \mathcal{U}$ where $\mathcal{L} = \{\mathbf{D}_i = (y_i, \mathbf{x}_i^\intercal)^\intercal : i = 1, \ldots, n\}$ are $n$ independent and identically distributed (iid) realizations from $p(y, \mathbf{x})$ and $\mathcal{U} = \{\mathbf{x}_i : i = n+1, \ldots, N+n\}$ are $N$ iid realizations from $p(\mathbf{x})$. We assume that (i) $\mathcal{L} \perp \mathcal{U}$, (ii) $N >> n$ so that $n/N \to 0$ as $n \to \infty$ and (iii) that the labeling mechanism is independent of $\mathbf{D} = (y, \mathbf{x}^\intercal)^\intercal$ and hence the underlying data of $\mathcal{L}$ and $\mathcal{U}$ are generated from the same distribution. Assumption (iii) is equivalent to the missing completely at random (MCAR) assumption in the missing data literature (Rubin, 1976). Thus, one may view the SSL problem as a missing data problem. The main difference, however, is highlighted by assumption (ii) which implies that the probability of missingness in the outcome tends to 1 as $n \to \infty$. Though the missingness mechanism is typically implicit in the SSL literature (Wasserman and Lafferty, 2008), we provide further discussion of this issue in Section 6.

## 2.2. Formulation of the Classification Rule

Our aim is to efficiently assess the accuracy of a binary classification rule for $y$ based on $\mathbf{x}$. To this end, we assume that a risk score $\widehat{\boldsymbol{\theta}}^{\mathsf{T}}\mathbf{x}$ is obtained by fitting a generalized linear model (GLM)

$$P(y = 1 \mid \mathbf{x}) = g(\boldsymbol{\theta}^{\mathsf{T}}\mathbf{x}), \quad \boldsymbol{\theta} = (\alpha, \boldsymbol{\beta}^{\mathsf{T}})^{\mathsf{T}}, \quad \boldsymbol{\beta} = (\beta_1, ..., \beta_p)^{\mathsf{T}} \tag{2.1}$$

where $\boldsymbol{\theta}$ is an unknown vector of regression parameters and $g(\cdot)$ is a specified smooth distribution function. To obtain a parsimonious classification rule as well as stabilize estimation when $p$ is not small relative to $n$ in finite simple, we employ a regularized procedure to obtain an estimator $\widehat{\boldsymbol{\theta}}$ for $\boldsymbol{\theta}$. For ease of presentation, we focus on simple one-step type penalty functions previously considered in Zou et al. (2008) and Minnier et al. (2011), but note that our procedure can be easily modified to accommodate other types of penalty functions which lead to a $\widehat{\boldsymbol{\theta}}$ with desirable oracle properties. Specifically, we let $\widehat{\boldsymbol{\theta}}$ be the minimizer of the penalized objective function

$$\widehat{\mathscr{L}}(\boldsymbol{\theta}) = n^{-1}\sum_{i=1}^{n}\mathscr{L}(\boldsymbol{\theta}, \mathbf{D}_i) + \sum_{j=1}^{p} p'_{\lambda_n j}(|\widetilde{\beta}_j|)|\beta_j| \tag{2.2}$$

where $\mathscr{L}(\boldsymbol{\theta}, \mathbf{D}) = -\log[g(\boldsymbol{\theta}^{\mathsf{T}}\mathbf{x})^y\{1 - g(\boldsymbol{\theta}^{\mathsf{T}}\mathbf{x})\}^{1-y}]$, $p'_{\lambda_n j}(|\beta_j|)$ is the derivative of a penalty function $p_{\lambda_n j}(|\beta_j|)$, $\widetilde{\boldsymbol{\theta}} = (\widetilde{\alpha}, \widetilde{\beta}_1, ..., \widetilde{\beta}_p)^{\mathsf{T}}$ is an initial estimator obtained as the minimizer of $n^{-1}\sum_{i=1}^{n}\mathscr{L}(\boldsymbol{\theta}, \mathbf{D}_i) + \lambda_{2n}\boldsymbol{\beta}^{\mathsf{T}}\boldsymbol{\beta}$ for some small $\lambda_{2n} = o(n^{-1/2})$, and $\lambda_n$ is such that $\lambda_n n^{1/2} \to \infty$ and $\lambda_n \to 0$ as $n \to \infty$. Letting $\boldsymbol{\theta}_0 = \text{argmin}_{\boldsymbol{\theta}}\mathbb{P}\{\mathscr{L}(\boldsymbol{\theta}, \mathbf{D})\}$, it has been shown in Minnier et al. (2011) that under mild regularity conditions with properly chosen $p_{\lambda_n j}(|\beta_j|)$,

$$n^{\frac{1}{2}}(\widehat{\boldsymbol{\theta}}_{\mathcal{A}} - \boldsymbol{\theta}_{0\mathcal{A}}) = \mathbb{G}_n\{\mathbf{V}_{\boldsymbol{\theta}_{\mathcal{A}}}(\mathbf{D})\} + o_p(1), \; n^{\frac{1}{2}}\widehat{\boldsymbol{\theta}}_{\mathcal{A}^c} = o_p(1), \; \text{and } \mathbb{P}(\widehat{\boldsymbol{\theta}}_{\mathcal{A}^c} = 0) \to 1 \tag{2.3}$$

where $\mathbb{G}_n = n^{1/2}(\mathbb{P}_n - \mathbb{P})$, $\mathbb{P}$ and $\mathbb{P}_n$ respectively denote the underlying and empirical probability measures generated by $\mathcal{L}$, $\mathbf{V}_{\boldsymbol{\theta}_{\mathcal{A}}}(\mathbf{D}) = \mathbb{A}_{11}^{-1}\mathbf{U}_{\mathcal{A}}(\boldsymbol{\theta}_0, \mathbf{D})$, $\mathbf{U}(\boldsymbol{\theta}, \mathbf{D}) = \partial\mathscr{L}(\boldsymbol{\theta}, \mathbf{D})/\partial\boldsymbol{\theta}$, $\mathbb{A} = \partial^2\mathbb{P}\{\mathscr{L}(\boldsymbol{\theta}, \mathbf{D})\}/\partial\boldsymbol{\theta}\boldsymbol{\theta}^{\mathsf{T}}|_{\boldsymbol{\theta}=\boldsymbol{\theta}_0}$, $\mathcal{A} = \{j : \theta_{0j} \neq 0\}$, $\mathbf{u}_{\mathcal{A}}$ denotes the subvector of $\mathbf{u}$ corresponding to $\mathcal{A}$, and $\mathbb{A}_{11}$ is the $q \times q$ submatrix of $\mathbb{A}$ corresponding to $q$ elements in $\mathcal{A}$. Note that this class of estimators is fairly general and includes the SCAD (Fan and Li, 2001; Zou et al., 2008), the adaptive LASSO (ALASSO)(Zou, 2006; Wang and Leng, 2007) and the standard GLM with the penalty parameters set to 0. We consider this relatively standard estimation procedure for $\boldsymbol{\theta}_0$ as our interest lies in using $\mathcal{U}$ to improve the estimation of prediction performance metrics rather than prediction performance itself.

## 2.3. Quantities of Interest

With a given $\widehat{\boldsymbol{\theta}}$, we may classify all future subjects with predicted risk scores in the highest percentiles as having the phenotype of interest. To identify a desirable threshold for the percentile and evaluate the classification accuracy, we consider the commonly employed ROC analysis (Pepe, 2003) of the risk percentile $\mathcal{P}^0_{\widehat{\boldsymbol{\theta}}} \equiv \mathcal{P}_{\widehat{\boldsymbol{\theta}}}(\mathbf{x}^0) \equiv F_{\widehat{\boldsymbol{\theta}}}(\widehat{\boldsymbol{\theta}}^{\mathsf{T}}\mathbf{x}^0)$ for the true phenotype $y^0$ where $F_{\boldsymbol{\theta}}(x) = P(\boldsymbol{\theta}^{\mathsf{T}}\mathbf{x} \leq x)$. Specifically, let

$$\overline{\text{TPR}}(c) = P(\mathcal{P}^0_{\widehat{\boldsymbol{\theta}}} \geq c | y^0 = 1) \quad \text{and} \quad \overline{\text{FPR}}(c) = P(\mathcal{P}^0_{\widehat{\boldsymbol{\theta}}} \geq c | y^0 = 0)$$

denote the expected true positive rate (TPR) and false positive rate (FPR) functions of $\mathcal{P}^0_{\widehat{\boldsymbol{\theta}}}$, where the probability is taken over the distribution of $\mathcal{L}$ and $\mathbf{D}^0 = (y^0, \mathbf{x}^{0\mathsf{T}})^{\mathsf{T}}$. These parameters quantify the expected prediction performance of $\mathcal{P}^0_{\widehat{\boldsymbol{\theta}}}$ averaged over the distribution of $\mathcal{L}$ at a given sample size of $n$ as we seek to evaluate the accuracy of the classification rule derived from the working regression model estimated with the available labeled data. One may summarize the trade-off between the TPR and FPR functions based on the ROC curve,

$$\overline{\text{ROC}}(u) = \overline{\text{TPR}}\left\{\overline{\text{FPR}}^{-1}(u)\right\}.$$

A threshold value $c_0$ for classifying a patient as having the phenotype, namely $\mathcal{P}^0_{\widehat{\boldsymbol{\theta}}} \geq c_0$, is often chosen to achieve a desired FPR level $u_0$, particularly when the prevalence of $y$ is low (Baker, 2003). Additionally, the area under the ROC curve, AUC $= \int_0^1 \overline{\text{ROC}}(u)du$, summarizes the overall prediction performance of $\mathcal{P}^0_{\widehat{\boldsymbol{\theta}}}$.

Once a threshold value $c_0$ is determined, it is necessary to evaluate the predictive performance of the binary rule $I(\mathcal{P}^0_{\widehat{\boldsymbol{\theta}}} \geq c_0)$, frequently summarized based on the positive predictive value (PPV) and negative predictive value (NPV) defined as

$$\overline{\text{PPV}}(c_0) = P(y^0 = 1 \mid \mathcal{P}^0_{\widehat{\boldsymbol{\theta}}} \geq c_0) \quad \text{and} \quad \overline{\text{NPV}}(c_0) = P(y^0 = 0 \mid \mathcal{P}^0_{\widehat{\boldsymbol{\theta}}} < c_0)$$

In evaluating EMR-based phenotyping algorithms, for example, the PPV measures the rate of concordance between positive classifications and true disease. A high PPV is therefore desirable as accurate prediction of disease status is a prerequisite for use of an algorithm in practice (Liao et al., 2015). We next detail our proposed SS estimators of these accuracy parameters and demonstrate that they are more efficient than their supervised counterparts.

# 3. SS Estimation of Classification Performance

## 3.1. Estimation

To motivate the SS estimator, we first note that the supervised estimators of $\overline{\text{TPR}}(c)$ and $\overline{\text{FPR}}(c)$ can be respectively constructed based on $\mathcal{L}$ alone as

$$\widetilde{\text{TPR}}(c) = \frac{\sum_{i=1}^n I(\widetilde{\mathcal{P}}_{\widehat{\boldsymbol{\theta}}i} \geq c)y_i}{\sum_{i=1}^n y_i} \quad \text{and} \quad \widetilde{\text{FPR}} = \frac{\sum_{i=1}^n I(\widetilde{\mathcal{P}}_{\widehat{\boldsymbol{\theta}}i} \geq c)(1-y_i)}{\sum_{i=1}^n (1-y_i)} \tag{3.1}$$

where $\widetilde{\mathcal{P}}_{\boldsymbol{\theta}i} = \widetilde{\mathcal{P}}_{\boldsymbol{\theta}}(\mathbf{x}_i)$, $\widetilde{\mathcal{P}}_{\boldsymbol{\theta}}(\mathbf{x}) = \widetilde{F}_{\boldsymbol{\theta}}(\boldsymbol{\theta}^\mathsf{T}\mathbf{x})$, and $\widetilde{F}_{\boldsymbol{\theta}}(t) = n^{-1}\sum_{i=1}^n I(\boldsymbol{\theta}^\mathsf{T}\mathbf{x}_i \leq t)$ is the supervised estimator of $F_{\boldsymbol{\theta}}(t)$. These estimators are the so-called 'apparent' or 'resubstitution' estimators as they utilize $\mathcal{L}$ for the estimation of $\boldsymbol{\theta}_0$ as well as the corresponding accuracy measure. To obtain more efficient SS estimators of these quantities, we wish to make use of $\mathcal{U}$ in addition to $\mathcal{L}$. If the outcomes in $\mathcal{U}$ were actually observed, we would simply compute (3.1) over $\mathcal{U}$. In the absence of true phenotype information, we make use of $\mathcal{U}$ by noting that

$$\overline{\text{TPR}}(c) = \frac{E\{y^0 I(\mathcal{P}^0_{\bar{\boldsymbol{\theta}}} \geq c)\}}{E(y^0)} = \frac{E\{\bar{m}(\mathcal{P}^0_{\bar{\boldsymbol{\theta}}})I(\mathcal{P}^0_{\bar{\boldsymbol{\theta}}} \geq c)\}}{E\{\bar{m}(\mathcal{P}_{\widehat{\boldsymbol{\theta}}^0})\}} \tag{3.2}$$

where $\bar{m}(s) = m(s, \mathcal{P}^0_{\bar{\boldsymbol{\theta}}})$, $m(s, \mathcal{P}_{\boldsymbol{\theta}}) = P\{y = 1 | \mathcal{P}_{\boldsymbol{\theta}}(\mathbf{x}) = s\}$ and with a slight abuse of notation we let $\mathcal{P}_\theta = \mathcal{P}_\theta(\mathbf{x}) = F_{\boldsymbol{\theta}}(\boldsymbol{\theta}^\intercal\mathbf{x})$. Similar to the estimation of $\mu$, the expression in (3.2) highlights (i) the dependence of $\overline{\text{TPR}}(c)$ on the distribution of $\mathbf{x}$ and hence the potential utility of $\mathcal{U}$ in improving estimation precision and (ii) the suitability of "imputing" the missing phenotype in $\mathcal{U}$ using $\bar{m}(\cdot)$ estimated from $\mathcal{L}$. We therefore propose the following two-step SS estimation procedures for the classification accuracy parameters.

In step I, we obtain a non-parametrically calibrated estimate of the conditional risk function $\bar{m}(s)$ via kernel smoothing as $\widetilde{m}(s, \widehat{\mathcal{P}}_{\widehat{\boldsymbol{\theta}}})$ where

$$\widetilde{m}(s, \mathcal{P}) = \frac{\sum_{i=1}^n K_h\{\mathcal{P}(\mathbf{x}_i) - s\}y_i}{\sum_{i=1}^n K_h\{\mathcal{P}(\mathbf{x}_i) - s\}} \tag{3.3}$$

$\widehat{\mathcal{P}}_{\boldsymbol{\theta}} \equiv \widehat{\mathcal{P}}_{\boldsymbol{\theta}}(\mathbf{x}) = \widehat{F}_{\boldsymbol{\theta}}(\boldsymbol{\theta}^\intercal\mathbf{x})$, $\widehat{F}_{\boldsymbol{\theta}}(t) = N^{-1}\sum_{i=n+1}^{N+n} I(\boldsymbol{\theta}^\intercal\mathbf{x}_i \leq t)$ is the SS estimate of $F_{\boldsymbol{\theta}}(t)$ based on $\mathcal{U}$, $K_h(u) = h^{-1}K(u/h)$, $K(\cdot)$ is a given smooth symmetric kernel density function with bounded support, and $h = h(n)$ is a bandwidth such that $nh^2 \to \infty$ and $nh^4 \to 0$ as $n \to \infty$. For ease of notation, we let $\widetilde{m}(s, \widehat{\mathcal{P}}_{\widehat{\boldsymbol{\theta}}}) = \widetilde{m}(s)$ throughout. In step II, we use $\widetilde{m}(\cdot)$ along with the data in $\mathcal{U}$ to construct the SS estimator of $\overline{\text{TPR}}(c)$,

$$\widehat{\text{TPR}}(c) = \frac{\sum_{i=n+1}^{N+n} I(\widehat{\mathcal{P}}_{\widehat{\boldsymbol{\theta}}i} \geq c)\widetilde{m}(\widehat{\mathcal{P}}_{\widehat{\boldsymbol{\theta}}i})}{\sum_{i=n+1}^{N+n}\widetilde{m}(\widehat{\mathcal{P}}_{\widehat{\boldsymbol{\theta}}i})}, \quad \text{where} \quad \widehat{\mathcal{P}}_{\boldsymbol{\theta}i} = \widehat{\mathcal{P}}_{\boldsymbol{\theta}}(\mathbf{x_i}).$$

This semi-nonparametric approach to imputation is particularly appealing as it only requires one-dimensional smoothing and protects against misspecification of the fitted regression model (2.1). That is, the calibrated estimator $\widetilde{m}(s)$ consistently estimates $\bar{m}(s)$ regardless of the adequacy of the fitted model (2.1). As a result, this procedure allows for valid inference about the prediction performance of $\mathcal{P}^0_{\bar{\boldsymbol{\theta}}}$ without requiring (2.1) to hold. It is also important to note that although $\widetilde{m}(\cdot)$ converges at a slower nonparametric rate, $\widehat{\text{TPR}}(c)$ still converges at a rate of root-$n$ as it is essentially integrated over the distribution of $\widehat{\mathcal{P}}_{\widehat{\boldsymbol{\theta}}i}$. The conditions on $h$ ensure undersmoothing to overcome the bias-variance trade-off.

Similarly, we may construct a SS estimator of $\overline{\mathrm{FPR}}(c)$ as

$$\widehat{\mathrm{FPR}}(c) = \frac{\sum_{i=n+1}^{N+n} I(\widehat{\mathcal{P}}_{\widehat{\boldsymbol{\theta}}i} \geq c)\{1 - \widetilde{m}(\widehat{\mathcal{P}}_{\widehat{\boldsymbol{\theta}}i})\}}{\sum_{i=n+1}^{N+n}\{1 - \widetilde{m}(\widehat{\mathcal{P}}_{\widehat{\boldsymbol{\theta}}i})\}};$$

while $\overline{\mathrm{PPV}}(c)$ and $\overline{\mathrm{NPV}}(c)$ may be consistently estimated using $\widehat{\mathrm{FPR}}(c)$ and $\widehat{\mathrm{TPR}}(c)$ with

$$\widehat{\mathrm{PPV}}(c) = \frac{\widehat{\mathrm{TPR}}(c)\hat{\mu}}{\widehat{\mathrm{FPR}}(c)\hat{\mu}_0 + \widehat{\mathrm{TPR}}(c)\hat{\mu}} \quad \text{and} \quad \widehat{\mathrm{NPV}}(c) = \frac{\{1 - \widehat{\mathrm{FPR}}(c)\}\hat{\mu}_0}{\{1 - \widehat{\mathrm{FPR}}(c)\}\hat{\mu}_0 + \{1 - \widehat{\mathrm{TPR}}(c)\}\hat{\mu}}$$

where $\hat{\mu} = N^{-1}\sum_{i=n+1}^{N+n} \widetilde{m}(\widehat{\mathcal{P}}_{\widehat{\boldsymbol{\theta}}i})$ is the SS estimator of the prevalence $\mu = P(y = 1)$ and $\hat{\mu}_0 = 1 - \hat{\mu}$. When a threshold value is selected as $\bar{c}_{u_0} = \overline{\mathrm{FPR}}^{-1}(u_0)$, we may obtain SS estimators of $\bar{c}_{u_0}$ as $\widehat{c}_{u_0} = \widehat{\mathrm{FPR}}^{-1}(u_0)$ and $\overline{\mathrm{ROC}}(u_0)$ as $\widehat{\mathrm{ROC}}(u_0) = \widehat{\mathrm{TPR}}(\widehat{c}_{u_0}) = \widehat{\mathrm{TPR}}\{\widehat{\mathrm{FPR}}^{-1}(u_0)\}$.

In the supervised setting, it is well-known that the apparent estimator in (3.1) may be overly optimistic in finite sample (Efron, 1986). A commonly used simple method to negate such overfitting bias is cross-validation (CV) which randomly splits $\mathcal{L}$ into mutually exclusive subsets to estimate the classifier and the accuracy parameters of interest. To reduce the overfitting bias in the proposed estimator, we develop here a $\mathcal{K}$-fold CV procedure in the SS setting. It is important to note that the apparent SS estimator is subject to overfitting as $\mathcal{L}$ is used to estimate both the risk score *and* the conditional risk function $\bar{m}(\cdot)$ which is utilized for estimation of the accuracy parameters. CV for the SS estimator correspondingly involves partitioning $\mathcal{L}$ for the estimation of $\boldsymbol{\theta}_0$ and $\bar{m}(\cdot)$ while the performance measure is estimated with all of $\mathcal{U}$ since $N$ is assumed to be sufficiently large. To this end, denote each fold of $\mathcal{L}$ as $\mathcal{I}_k$ and the corresponding indices as $\mathbb{I}_k$ for $k = 1, \ldots, \mathcal{K}$. For a given $k$, we fit the regression model with $\mathcal{L} \setminus \mathcal{I}_k$ to obtain an estimator for $\boldsymbol{\theta}_0$ denoted as $\widehat{\boldsymbol{\theta}}^{(\text{-}k)}$. The observations in the left-out set $\mathcal{I}_k$ are used to estimate $\bar{m}(s)$ with the local constant smoother as

$$\widetilde{m}_k(s) = \widetilde{m}_k(s, \widehat{\mathcal{P}}_{\widehat{\boldsymbol{\theta}}^{(\text{-}k)}}) = \frac{\sum_{i \in \mathbb{I}_k} K_h(\widehat{\mathcal{P}}_{\widehat{\boldsymbol{\theta}}^{(\text{-}k)}i} - s)y_i}{\sum_{i \in \mathbb{I}_k} K_h(\widehat{\mathcal{P}}_{\widehat{\boldsymbol{\theta}}^{(\text{-}k)}i} - s)}$$

where $\widehat{\mathcal{P}}_{\widehat{\boldsymbol{\theta}}^{(-k)}i} = \widehat{F}_{\widehat{\boldsymbol{\theta}}^{(-k)}}\{(\widehat{\boldsymbol{\theta}}^{(-k)})^{\intercal}\mathbf{x}_i\}$. We then estimate $\overline{\mathrm{TPR}}(c)$ as

$$\widehat{\mathrm{TPR}}_k(c) = \frac{\sum_{i=n+1}^{N+n} I(\widehat{\mathcal{P}}_{\widehat{\boldsymbol{\theta}}^{(-k)}i} \geq c)\widetilde{m}_k(\widehat{\mathcal{P}}_{\widehat{\boldsymbol{\theta}}^{(-k)}i})}{\sum_{i=n+1}^{N+n} \widetilde{m}_k(\widehat{\mathcal{P}}_{\widehat{\boldsymbol{\theta}}^{(-k)}i})}$$

and the final CV estimator for $\overline{\mathrm{TPR}}(c)$ is $\widehat{\mathrm{TPR}}_{\mathrm{cv}}(c) = \mathcal{K}^{-1}\sum_{k=1}^{\mathcal{K}} \widehat{\mathrm{TPR}}_k(c)$. In practice, we suggest averaging over several repetitions of CV to minimize the additional variation induced by random partitioning. Similarly, we may construct CV estimators for $\overline{\mathrm{FPR}}(c)$, $\overline{\mathrm{ROC}}(c)$, $\overline{\mathrm{PPV}}(c)$, and $\overline{\mathrm{NPV}}(c)$, respectively denoted as $\widehat{\mathrm{FPR}}_{\mathrm{cv}}(c)$, $\widehat{\mathrm{ROC}}_{\mathrm{cv}}(c)$, $\widehat{\mathrm{PPV}}_{\mathrm{cv}}(c)$, and $\widehat{\mathrm{NPV}}_{\mathrm{cv}}(c)$.

## 3.2. Asymptotic Results for the SS estimators

Though analogous results hold for all of the SS estimators, we present the main result for $\widehat{\mathrm{ROC}}(u_0)$ and demonstrate that it is asymptotically more efficient than the supervised estimator of $\overline{\mathrm{ROC}}(u_0)$, $\widetilde{\mathrm{ROC}}(u_0) = \widetilde{\mathrm{TPR}}\{\widetilde{\mathrm{FPR}}^{-1}(u_0)\}$. Throughout, we let $\mathcal{P}_{\boldsymbol{\theta}}(\mathbf{x}) = F_{\boldsymbol{\theta}}(\boldsymbol{\theta}^{\intercal}\mathbf{x})$ and $\mathrm{FPR}(c)$, $\mathrm{TPR}(c)$, and $\mathrm{ROC}(c)$ denote the population versions of TPR, FPR, and ROC functions for $\mathcal{P}^0_{\boldsymbol{\theta}_0}$.

**Theorem 1.** *Under the assumptions given in the Appendix in the Supplementary Materials,* $\widehat{\mathcal{W}}_{ROC}(u_0) = n^{\frac{1}{2}}\{\widehat{ROC}(u_0) - ROC(u_0)\} = \mathbb{G}_n\left\{\mathcal{W}^{SS}_{ROC}(\boldsymbol{\theta}_0, u_0, \mathbf{D})\right\} + o_p(1)$ *and* $\widetilde{\mathcal{W}}_{ROC}(u_0) = n^{1/2}\{\widetilde{ROC}(u_0) - ROC(u_0)\} = \mathbb{G}_n\{\mathcal{W}^{SL}_{ROC}(\boldsymbol{\theta}_0, u_0, \mathbf{D})\} + o_p(1)$, *where*

$$\mathcal{W}^{SS}_{ROC}(\boldsymbol{\theta}_0, u_0, \mathbf{D}_i) = \mathcal{G}_{u_0}(\mathcal{P}_{\boldsymbol{\theta}_0 i})\{y_i - E(y_i \mid \mathcal{P}_{\boldsymbol{\theta}_0 i})\} - \mathcal{J}_{u_0}(\mathbf{D}_i), \tag{3.4}$$

$$\mathcal{W}^{SL}_{ROC}(\boldsymbol{\theta}_0, u_0, \mathbf{D}_i) = \mathcal{G}_{u_0}(\mathcal{P}_{\boldsymbol{\theta}_0 i})(y_i - \mu) - \mathcal{J}_{u_0}(\mathbf{D}_i). \tag{3.5}$$

$\mathcal{G}_{u_0}(\mathcal{P}_{\boldsymbol{\theta}_0 i}) = (\mu^{-1} + \kappa_{u_0})I(\mathcal{P}_{\boldsymbol{\theta}_0 i} \geq c_{u_0}) - \gamma_{u_0}$, $\kappa_{u_0} = \mu_0^{-1}\dot{ROC}(u_0)$, $\dot{ROC}(u_0) = \frac{dROC(u)}{du}\Big|_{u=u_0}$, $\gamma_{u_0} = \mu^{-1}ROC(u_0) + \kappa_{u_0}u_0$, $\mathcal{J}_{u_0}(\mathbf{D}_i) = \{\mu^{-1}\boldsymbol{\psi}_{\mathcal{A}}(\boldsymbol{\theta}_0, c_{u_0}) + \kappa_{u_0}\boldsymbol{\phi}_{\mathcal{A}}(\boldsymbol{\theta}_0, c_{u_0})\}^{\intercal}\mathbf{V}_{\boldsymbol{\theta}_{\mathcal{A}}}(\mathbf{D}_i)$, *and* $\boldsymbol{\psi}_{\mathcal{A}}(\boldsymbol{\theta}, c)$ *and* $\boldsymbol{\phi}_{\mathcal{A}}(\boldsymbol{\theta}, c)$ *are defined in the Appendix.*

Roughly speaking, the first term of (3.4) and (3.5) accounts for the variation in the accuracy measure while the second (and equivalent) term accounts for the variability in $\widehat{\boldsymbol{\theta}}$. Momentarily focusing on the first term, which provides the improvement of the SS approach, we note the similarity to the expansions for $\mu$. That is, the influence function for the SS estimator is centered at the conditional mean $E(y_i \mid \mathcal{P}_{\boldsymbol{\theta}_0 i})$ while its supervised counterpart is centered at the marginal mean $\mu = P(y = 1)$ thereby yielding a reduction in the asymptotic variance. More formally, following these expansions, it is straightforward to show that

$$\Delta_v(u_0) \equiv n \left[ \text{var}\left\{ \widetilde{\text{ROC}}(u_0) \right\} - \text{var}\left\{ \widehat{\text{ROC}}(u_0) \right\} \right] \approx$$
$$\text{var}\left[ E\{\mathcal{J}_{u_0}(\mathbf{D}_i) \mid \mathcal{P}_{\boldsymbol{\theta}_0 i}\} - \mathcal{G}_{u_0}(\mathcal{P}_{\boldsymbol{\theta}_0 i})\{E(y_i \mid \mathcal{P}_{\boldsymbol{\theta}_0 i}) - \mu\} \right] - \text{var}\left[ E\{\mathcal{J}_{u_0}(\mathbf{D}_i) \mid \mathcal{P}_{\boldsymbol{\theta}_0 i}\} \right]. \qquad (3.6)$$

Provided that $E\{\mathcal{J}_{u_0}(\mathbf{D}_i) \mid \mathcal{P}_{\boldsymbol{\theta}_0 i}\} = 0$, which holds under correct model specification, $\Delta_v(u_0) \approx \text{var}[\mathcal{G}_{u_0}(\mathcal{P}_{\boldsymbol{\theta}_0 i})\{E(y_i \mid \mathcal{P}_{\boldsymbol{\theta}_0 i}) - \mu\}]$ and the SS estimator of $\widehat{\text{ROC}}(u_0)$ is *always* asymptotically more efficient than its supervised counterpart when there is an association between $y$ and $\mathbf{x}$. Under slight model mis-specification, one would still expect the SS estimator to be more efficient since $E\{\mathcal{J}_{u_0}(\mathbf{D}_i) \mid \mathcal{P}_{\boldsymbol{\theta}_0 i}\}$ is typically small in magnitude and $\Delta_v(u_0)$ is dominated by the term $\text{var}[\mathcal{G}_{u_0}(\mathcal{P}_{\boldsymbol{\theta}_0 i})\{E(y_i \mid \mathcal{P}_{\boldsymbol{\theta}_0 i}) - \mu\}]$. Our simulation results in Section 4 support this claim.

For the CV estimators, we show in the Appendix that $\widehat{\mathcal{W}}_{\text{ROC}}(u_0)$ and $n^{1/2}\{\widehat{\text{ROC}}_{\text{cv}}(u_0) - \text{ROC}(u_0)\}$ are asymptotically equivalent with the same limiting distribution. Similar findings have been noted in Tian et al. (2007) for absolute prediction error estimators when no regularization or smoothing was employed for the estimation. Although this result suggests that one may approximate the standard error (SE) of $\widehat{\text{ROC}}_{\text{cv}}(u_0)$ based on the SE estimate of $\widehat{\text{ROC}}(u_0)$, we find that such an approximation does not perform well when $p$ is not very small relative to $n$ due to the overfitting bias. We next propose CV-based perturbation resampling procedures that provide more accurate SE estimates by correcting for overfitting.

## 3.3. Perturbation Resampling Procedure for Inference

To make inference about $\overline{\text{ROC}}(u_0)$ based on $\widehat{\text{ROC}}_{\text{cv}}(u_0)$, we rely on the asymptotic normality of $\widehat{\mathcal{W}}_{\text{ROC}}(u_0)$. However, the influence function expansion in (3.4) reveals that the asymptotic variance of $\widehat{\text{ROC}}_{\text{cv}}(u_0)$ involves unknown conditional density functions which are difficult to estimate explicitly, particularly under model mis-specification. To overcome this difficulty as well as overfitting, we propose a hybrid of CV and perturbation resampling technique (Jin et al., 2001) to obtain variance estimates for $\widehat{\mathrm{ROC}}_{\mathrm{cv}}(u_0)$. To this end, we generate a set of iid non-negative random variables, $\mathcal{G} = (G_1, \dots, G_n)^{\mathsf{T}}$, independent of $\mathcal{D}$, following a known distribution with mean one and unit variance.

We first obtain a resampled counterpart of $\widehat{\boldsymbol{\theta}}$ by perturbing a bias corrected estimate of the influence function given in (2.3). Specifically, let $\widehat{\boldsymbol{\theta}}^*_{\widehat{\mathcal{A}}^c} = 0$ and

$$\widehat{\boldsymbol{\theta}}^*_{\widehat{\mathcal{A}}} = \widehat{\boldsymbol{\theta}}_{\widehat{\mathcal{A}}} - n^{-1} \sum_{k=1}^{\mathcal{K}} \sum_{i \in \mathbb{I}_k} \widehat{\mathbb{A}}_{11}^{-1} \mathbf{U}_{\widehat{\mathcal{A}}}(\widehat{\boldsymbol{\theta}}_{(-k)}, \mathbf{D}_i)(G_i - 1)$$

where $\widehat{\mathbb{A}}$ is the empirical estimate of $\mathbb{A}$ and $\widehat{\mathbb{A}}_{11}$ is the submatrix of $\widehat{\mathbb{A}}$ corresponding to $\widehat{\mathcal{A}} = \{l : \widehat{\theta}_l \neq 0\}$. To account for the variation in $\widetilde{m}(s)$ and correct for overfitting, we note that

$$\widetilde{m}(s) - m(s, \mathcal{P}_{\boldsymbol{\theta}_0}) = \widetilde{m}(s, \mathcal{P}_{\boldsymbol{\theta}_0}) - m(s, \mathcal{P}_{\boldsymbol{\theta}_0}) + \widetilde{m}(s, \widehat{\mathcal{P}}_{\widehat{\boldsymbol{\theta}}}) - \widetilde{m}(s, \mathcal{P}_{\boldsymbol{\theta}_0}).$$

We construct a perturbed counterpart of $\widetilde{m}(s)$ as $\widetilde{m}^*(s) = \widetilde{m}^*_A(s) + \widetilde{m}(s, \widehat{\mathcal{P}}_{\widehat{\boldsymbol{\theta}}^*})$ where

$$\widetilde{m}^*_A(s) = \frac{\sum_{k=1}^{\mathcal{K}} \sum_{i \in \mathbb{I}_k} K_h(\widehat{\mathcal{P}}_{\widehat{\boldsymbol{\theta}} i} - s)\{y_i - \widetilde{m}_{(\text{-}k)}(\widehat{\mathcal{P}}_{\widehat{\boldsymbol{\theta}}^{(\text{-}k)} i})\} G_i}{\sum_{i=1}^n K_h(\widehat{\mathcal{P}}_{\widehat{\boldsymbol{\theta}} i} - s) G_i} \text{ and } \widetilde{m}_{(-k)}(s) = \frac{\sum_{i \in \mathbb{I}_k^c} K_h(\widehat{\mathcal{P}}_{\widehat{\boldsymbol{\theta}}^{(\text{-}k)} i} - s) y_i}{\sum_{i \in \mathbb{I}_k^c} K_h(\widehat{\mathcal{P}}_{\widehat{\boldsymbol{\theta}}^{(\text{-}k)} i} - s)}.$$

Note that $\widetilde{m}^*_A(s)$ accounts for the variation in $\widetilde{m}(s)$ attributable to the non-parametric smoothing ignoring the variation in $\widehat{\boldsymbol{\theta}}$ while $\widetilde{m}(s, \widehat{\mathcal{P}}_{\widehat{\boldsymbol{\theta}}^*})$ accounts for the variation in $\widehat{\boldsymbol{\theta}}$. Finally, we obtain a perturbed counterpart of $\widehat{\mathrm{TPR}}_{\mathrm{cv}}(c)$ as

$$\widehat{\mathrm{TPR}}^*(c) = \frac{\sum_{i=n+1}^{N+n} I(\widehat{\mathcal{P}}_{\widehat{\boldsymbol{\theta}}^* i} \geq c)\{\widetilde{m}^*_A(\widehat{\mathcal{P}}_{\widehat{\boldsymbol{\theta}} i}) + \widetilde{m}(\widehat{\mathcal{P}}_{\widehat{\boldsymbol{\theta}}^* i})\}}{\sum_{i=n+1}^{N+n} \{\widetilde{m}^*_A(\widehat{\mathcal{P}}_{\widehat{\boldsymbol{\theta}} i}) + \widetilde{m}(\widehat{\mathcal{P}}_{\widehat{\boldsymbol{\theta}}^* i})\}}.$$

We may obtain a perturbed counterpart of $\widehat{\mathrm{FPR}}_{\mathrm{cv}}(c)$ accordingly as $\widehat{\mathrm{FPR}}^*(c)$ and let $\widehat{\mathrm{ROC}}^*(c) = \widehat{\mathrm{TPR}}^*\{\widehat{\mathrm{FPR}}^{*-1}(u_0)\}$. The variance of $\widehat{\mathrm{ROC}}_{\mathrm{cv}}(u_0)$ can therefore be consistently estimated using the empirical distribution of $\widehat{\mathrm{ROC}}^*(c)$ and the corresponding interval estimates can be constructed according to the asymptotic normal distribution of the SS estimator. Confidence intervals for $\overline{\mathrm{ROC}}(u_0)$ are constructed by centering at $\widehat{\mathrm{ROC}}_{\mathrm{cv}}(u_0)$ with width determined by the empirical SE of $\widehat{\text{ROC}}^*(c)$.

# 4. Simulation Studies

We conducted extensive simulation studies to validate the proposed point and interval estimation procedures. Throughout, we generated $\mathbf{x}$ from $\text{MVN}(0,\ 3\rho + 3(1-\rho)\mathbb{I}_{p\times p}) + \text{Bin}(3, 0.3)\mathbf{1}_{p\times 1}$ with $\rho$ chosen to be either 0.2 or 0.4 and $p = 10$ or 20. To build a classifier for $y$, we fit (2.1) with $g(x) = \text{expit}(x)$ under three data generating mechanisms:

1. Correct specification: $y \sim \text{Bin}\{1, \text{expit}(\boldsymbol{\theta}_0^\intercal \mathbf{x})\}$;

2. Misspecified link function: $y \sim \text{Bin}\{1, \widetilde{g}(\boldsymbol{\theta}_0^\intercal \mathbf{x})\}$ with $\widetilde{g}(x) = \Phi\{(x+2)/2\}$;

3. Misspecified linear predictor: $y \sim \text{Bin}\{1, \text{expit}(\boldsymbol{\theta}_0^\intercal \mathbf{x} + x_3 x_4)\}$;

where $\boldsymbol{\theta}_0 = (-4, 1, 1, 0.5, 0.5, \mathbf{0}_{(p-4)\times 1}^\intercal)^\intercal$. We considered $n = 200$, $400$ and $N = 20,000$. The true values of the target parameters were estimated via Monte Carlo simulation with a large sample size of 50,000. For each configuration, results were summarized based on $1,500$ datasets.

Across all numerical studies including the data examples in Section 5, we used the ALASSO penalty with $p'_{\lambda_n j}(|\beta_j|) = n^{-1/2}\lambda_n|\beta_j|^{-1}$ to estimate $\boldsymbol{\theta}_0$, where we set $\lambda_{2n} = \log(p)/n^{1.5}$ and chose $\lambda_n$ using a modified BIC that replaces $\log(n)$ with $\min\{(\sum_{i=1}^n y_i)^{0.1}, \log(\sum_{i=1}^n y_i)\}$ to avoid excessive shrinkage in finite sample. For the non-parametric smoothing, we used the Gaussian kernel with $h = n^{-0.45}\hat{\sigma}_n$ where $\hat{\sigma}_n$ is the empirical standard deviation of $\{\widetilde{\mathcal{P}}_{\widehat{\boldsymbol{\theta}}i}\}_{i=1}^n$. To estimate the SE, we used 500 perturbations and employed a robust SE calculation which removes realizations more than 6 median absolute deviations (MAD) away from the median. This approximation is reasonable as the estimators are asymptotically normal and thus the probability of the realizations in the removed extreme tails is of order $10^{-9}$. For the CV procedures, we let $K = 10$ and averaged over 10 replications.

We present results for $\overline{\text{ROC}}(u_0)$, $\overline{\text{PPV}}(\bar{c}_{u_0})$, $\overline{\text{NPV}}(\bar{c}_{u_0})$ with $u_0 = 0.05$. In Table 1 we present the percent bias for the apparent and CV estimates of the accuracy measures under correct model specification. The results for settings 2 and 3 can be found in the Supplementary Material and follow similar patterns. As expected, the apparent estimators exhibit substantial bias in both the supervised and SS settings, particularly when $p = 20$ and $n = 200$. This is consistent with the general consensus of apparent estimators not being appropriate for prediction performance assessment. For the remaining evaluations, we therefore focus on the CV estimators only. Overall, the CV SS estimators are slightly less biased than the supervised estimators.

We summarize in Figure 1 the efficiency of the CV SS estimators relative to their supervised counterparts with respect to mean square error for all three data generating mechanisms. In each of these settings, the efficiency gains are significant for all parameters with gains as high as 174% for PPV and do not vary significantly with $\rho$, $n$, and $p$. We also note that the improvement in the estimation of the cut-off parameter $\bar{c}_{u_0}$ will directly impact the performance of the classifier. Additionally, the SS estimators outperform the supervised estimators under model misspecification with substantial efficiency gains in both settings 2 and 3.

The performance of the interval estimation based on the proposed perturbation resampling procedure is summarized in Table 2 under correct model specification. Overall, the resampling method is effective in standard error estimation with empirical standard errors close to the median of the estimated standard errors. The empirical coverage probabilities of the 95% confidence intervals are close to the nominal level across all settings. Results from settings 2 and 3 under model mis-specification, presented in the Supplementary Material, show similar patterns.

# 5. Application to EMR Studies

## 5.1. Background

The adoption of EMR in routine health care has resulted in a promising new data source for medical discovery research. Filled with comprehensive clinical information for extensive pa-tient populations, EMR can facilitate large-scale observational studies in a cost-effective and timely manner (Wilke et al., 2011). Additionally, when linked with specimen bio-repositories, these large medical databanks allow for the quantification of the effects of rare genetic variants as well as the study of complex genome-phenome architecture that can lead to discovery of new disease subtypes and their associated genetic causes (Kohane, 2011; Murphy et al., 2009). However, a major challenge in EMR-driven research is in the ascertainment of validated phenotype information as it requires substantial and thus prohibitive manual chart review. As a result, benchmark labels are only available for a small fraction of observations while predictors of phenotype are available for the entire cohort (Liao et al., 2010). This setting therefore directly lends itself to the use of SSL procedures.

## 5.2. Real Data Analysis

To illustrate our proposals, we applied our procedures to evaluate two phenotyping algorithms for classifying two systematic autoimmune disease conditions, RA and MS, using EMR data from Partner's HealthCare (Liao et al., 2010; Xia et al., 2013). For the RA study, the patient cohort consisted of $n = 500$ patients whose RA status was confirmed with medical chart review by two rheumatologists and a large 'RA mart' of $N = 43{,}514$ patients without confirmed disease status. Both narrative and codified data were available to develop the prediction model ($p = 37$). The narrative variables were obtained with natural language processing via the Health Information Text Extraction (HITex) system. These variables included disease diagnoses, medications, and radiology findings. The codified data included ICD9 codes, electronic prescriptions, and laboratory values. For the MS study, a neurologist confirmed MS status in $n = 455$ patients with at least one ICD9 code of MS via documentation in a neurologist's clinical note or a relevant MRI report in the medical records. An unlabeled data of size $N = 11{,}743$ was also available for analysis. We used codified variables ($p = 10$), including race, sex, gender, number of cervical spine and brain MRI, and ICD9 codes relating to MS, to develop the algorithm.

The estimates of prediction accuracy for $u_0 = 0.05$ as well as their estimated SEs for both studies are presented in Table 2. The point estimates of the parameters based on the supervised and SS methods are similar. This is a desirable property as it suggests the stability of the proposed procedure in a real data setting. We also observe substantial efficiency gains from the SS approach for each parameter in both studies. As a result, we have a more precise estimate of the prediction performance of the phenotyping algorithm using the SS method, suggesting the usefulness of our method in practice. For the RA study, efficiency gains are at least 85% for each accuracy measure with gains as high as 244% for the threshold parameter. For the MS study, the SS estimators are at least 60% more efficient than the supervised estimators and the SS estimator of PPV is 3.2 times more efficient than the supervised estimator. Importantly, the threshold parameter, which is ultimately used to determine classifications, is over 2.6 times more efficient than the corresponding supervised estimator.

## 6. Discussion

Unlike previous work in SSL, we have proposed a two-step estimation procedure that utilizes unlabeled data for model evaluation rather than model fitting. In particular, we introduced SS estimators of various prediction performance measures. Asymptotic results confirm that these estimators are always more efficient than their supervised counterparts under correct model specification. We addressed potential overfitting bias in our SS estimators with CV and also developed a CV-based perturbation resampling procedure that adjusts for sources of finite sample bias. Further, the SS estimator outperformed the supervised estimator in terms of efficiency in simulations as well as a real data analysis of two EMR-based studies thus illustrating the utility of our method in practice.

It is interesting to note that if model (2.1) is correctly specified, $\mathcal{P}_{\boldsymbol{\theta}_0}$ achieves the highest ROC curve among all scores based on $\mathbf{x}$ for the classification of $y$ (McIntosh and Pepe, 2002). Thus, under correct model specification, $\widehat{\boldsymbol{\theta}}$ does not contribute to the variability of $\widehat{\mathrm{ROC}}(u_0)$ asymptotically and the proposed SS estimator achieves the highest possible efficiency gain from $\mathcal{U}$. As it is unlikely that the working model is correctly specified, it would be of interest to potentially employ a 'safe' SS procedure to estimate $\boldsymbol{\theta}_0$ to provide further gains under model misspecification. Additionally, and perhaps a limitation of our study, is the typical SSL assumption of MCAR. Further work is needed to extend our results to the missing at random (MAR) setting (Rubin, 1976) to allow the labeling process to depend on $\mathbf{x}$.

Throughout, we focused on the setting with fixed $p$ but accommodated settings in which $p$ is not small relative to $n$ in finite sample with regularized estimation. For such settings, regularization has the advantage of producing more stable estimators for the model parameters and leading to more stable inference, compared to fitting the standard GLM. This is confirmed by results from a numerical study, shown in the Supplementary Material, comparing the performance of the proposed point and interval estimation procedures with $\widehat{\boldsymbol{\theta}}$ obtained from standard maximum likelihood and ALASSO in the setting with $p = 20, \rho = 0.4$, and $n = 200$ for the both the supervised and SS methods. These results indicate that (i) better performance of the prediction model is attained using penalized fitting versus a standard unpenalized fitting; (ii) the bias of the corresponding accuracy measures is significantly lower for the penalized approach; and (iii) unpenalized fitting leads to difficulties in constructing confidence intervals for the accuracy parameters with desired coverage levels, in contrast to those from penalized fitting. Lastly, while the theoretical results could be extended to allow $p$ growing with $n$ at a slow rate, SSL estimation under the setting with $p \gg n$ would require different theoretical justifications and warrant additional research.

# REFERENCES

Ananthakrishnan, A., Cai, T., Savova, G., Cheng, S., Chen, P., Perez, R., Gainer, V., Murphy, S., Szolovits, P., Xia, Z. et al. (2013) Improving case definition of crohn's disease and ulcerative colitis in electronic medical records using natural language processing: a novel informatics approach. *Inflammatory bowel diseases*, **19**, 1411–1420.

Baker, S. G. (2003) The central role of receiver operating characteristic (roc) curves in evaluating tests for the early detection of cancer. *Journal of the National Cancer Institute*, **95**, 511–515.

Baluja, S. (1998) Probabilistic modeling for face orientation discrimination learning from labeled and unlabeled data.

Belkin, M. and Niyogi, P. (2004) Semi-supervised learning on riemannian manifolds. *Machine learning*, **56**, 209–239.

Belkin, M., Niyogi, P. and Sindhwani, V. (2006) Manifold regularization: A geometric framework for learning from labeled and unlabeled examples. *The Journal of Machine Learning Research*, **7**, 2399–2434.

Bickel, P. J. and Rosenblatt, M. (1973) On some global measures of the deviations of density function estimates. *The Annals of Statistics*, 1071–1095.

Chapelle, O., Schölkopf, B., Zien, A. et al. (2006) *Semi-supervised learning*, vol. 2. MIT press Cambridge.

Cheng, Y., Zhao, X., Huang, K. and Tan, T. (2015) Semi-supervised learning and feature evaluation for rgb-d object recognition. *Computer Vision and Image Understanding*, **139**, 149–160.

Claesen, M., Davis, J., De Smet, F. and De Moor, B. (2015) *Assessing binary classifiers using only positive and unlabeled data.* arXiv preprint arXiv:1504.06837.

Collobert, R. and Weston, J. (2008) A unified architecture for natural language processing: Deep neural networks with multitask learning. In *Proceedings of the 25th international conference on Machine learning*, 160–167. ACM.

Cozman, F. G., Cohen, I. and Cirelo, M. (2002) Unlabeled data can degrade classification performance of generative classifiers. In *FLAIRS Conference*, 327–331.

Cozman, F. G., Cohen, I., Cirelo, M. C. et al. (2003) Semi-supervised learning of mixture models. In *ICML*, 99–106.

Efron, B. (1986) How biased is the apparent error rate of a prediction rule? *Journal of the American Statistical Association*, **81**, 461–470.

Fan, J. and Li, R. (2001) Variable selection via nonconcave penalized likelihood and its oracle properties. *Journal of the American statistical Association*, **96**, 1348–1360.

Grandvalet, Y. and Bengio, Y. (2004) Semi-supervised learning by entropy minimization. In *Advances in neural information processing systems*, 529–536.

Jaakkola, T., Haussler, D. et al. (1999) Exploiting generative models in discriminative classifiers. *Advances in neural information processing systems*, 487–493.

Jin, Z., Ying, Z. and Wei, L.-J. (2001) A simple resampling method by perturbing the minimand. *Biometrika*, **88**, 381–390.

Kohane, I. S. (2011) Using electronic health records to drive discovery in disease genomics. *Nature Reviews Genetics*, **12**, 417–428.

Lafferty, J. D. and Wasserman, L. (2006) Challenges in statistical machine learning.

Liao, K., Kurreeman, F., Li, G., Duclos, G., Murphy, S., RG, P., Cai, T., Gupta, N., Gainer, V., Schur, P. et al. (2012) Autoantibodies, autoimmune risk alleles and clinical associations in rheumatoid arthritis cases and non-ra controls in the electronic medical records. *Arthritis and rheumatism*.

Liao, K. P., Cai, T., Gainer, V., Goryachev, S., Zeng-treitler, Q., Raychaudhuri, S., Szolovits, P., Churchill, S., Murphy, S., Kohane, I. et al. (2010) Electronic medical records for discovery research in rheumatoid arthritis. *Arthritis care & research*, **62**, 1120–1127.

Liao, K. P., Cai, T., Savova, G. K., Murphy, S. N., Karlson, E. W., Ananthakrishnan, A. N., Gainer, V. S., Shaw, S. Y., Xia, Z., Szolovits, P. et al. (2015) Development of phenotype algorithms using electronic medical records and incorporating natural language processing. *bmj*, **350**, h1885.

Liu, R., Zhou, J. and Liu, M. (2006) Graph-based semi-supervised learning algorithm for web page classification. In *Intelligent Systems Design and Applications, 2006. ISDA'06. Sixth International Conference on*, vol. 2, 856–860. IEEE.

McIntosh, M. W. and Pepe, M. S. (2002) Combining several screening tests: optimality of the risk score. *Biometrics*, **58**, 657–664.

Minnier, J., Tian, L. and Cai, T. (2011) A perturbation method for inference on regularized regression estimates. *Journal of the American Statistical Association*, **106**, 1371–1382.

Murphy, S., Churchill, S., Bry, L., Chueh, H., Weiss, S., Lazarus, R., Zeng, Q., Dubey, A., Gainer, V., Mendis, M. et al. (2009) Instrumenting the health care enterprise for discovery research in the genomic era. *Genome research*, **19**, 1675–1681.

Nigam, K., McCallum, A. and Mitchell, T. (2006) Semi-supervised text classification using em. *Semi-Supervised Learning*, 33–56.

Niyogi, P. (2008) Manifold regularization and semi-supervised learning: Some theoretical analyses.

O'neill, T. J. (1978) Normal discrimination with unclassified observations. *Journal of the American Statistical Association*, **73**, 821–826.

Pepe, M. S. (2003) *The statistical evaluation of medical tests for classification and prediction.* Oxford University Press.

Rosenberg, C., Hebert, M. and Schneiderman, H. (2005) Semi-supervised self-training of object detection models. In *Application of Computer Vision, 2005. WACV/MOTIONS'05 Volume 1. Seventh IEEE Workshops on*, vol. 1, 29–36. IEEE.

Rubin, D. B. (1976) Inference and missing data. *Biometrika*, **63**, 581–592.

Seeger, M. (2000) Learning with labeled and unlabeled data. *Tech. rep.*, University of Edinburgh, United Kingdom.

Singh, A., Nowak, R. and Zhu, X. (2009) Unlabeled data: Now it helps, now it doesn't. In *Advances in neural information processing systems*, 1513–1520.

Sinnott, J. A., Dai, W., Liao, K. P., Shaw, S. Y., Ananthakrishnan, A. N., Gainer, V. S., Karlson, E. W., Churchill, S., Szolovits, P., Murphy, S. et al. (2014) Improving the power of genetic association tests with imperfect phenotype derived from electronic medical records. *Human genetics*, **133**, 1369–1382.

Søgaard, A. (2013) Semi-supervised learning and domain adaptation in natural language processing. *Synthesis Lectures on Human Language Technologies*, **6**, 1–103.

Tian, L., Cai, T., Goetghebeur, E. and Wei, L. (2007) Model evaluation based on the sampling distribution of estimated absolute prediction error. *Biometrika*, **94**, 297–311.

Wang, H. and Leng, C. (2007) Unified lasso estimation by least squares approximation. *Journal of the American Statistical Association*, **102**, 1039–1048.

Wang, Z. and Chen, S. (2011) Web page classification based on semi-supervised naïve bayesian em algorithm. In *Communication Software and Networks (ICCSN), 2011 IEEE 3rd International Conference on*, 242–245. IEEE.

Wasserman, L. and Lafferty, J. D. (2008) Statistical analysis of semi-supervised regression. In *Advances in Neural Information Processing Systems*, 801–808.

Wilke, R., Xu, H., Denny, J., Roden, D., Krauss, R., McCarty, C., Davis, R., Skaar, T., Lamba, J. and Savova, G. (2011) The emerging role of electronic medical records in pharmacogenomics. *Clinical Pharmacology & Therapeutics*, **89**, 379–386.

Xia, Z., Secor, E., Chibnik, L. B., Bove, R. M., Cheng, S., Chitnis, T., Cagan, A., Gainer, V. S., Chen, P. J., Liao, K. P. et al. (2013) Modeling disease severity in multiple sclerosis using electronic health records. *PloS one*, **8**, e78927.

Zhang, T. and Oles, F. (2000) The value of unlabeled data for classification problems. In *Proceedings of the Seventeenth International Conference on Machine Learning,(Langley, P., ed.)*, 1191–1198. Citeseer.

Zhu, X. (2006) Semi-supervised learning literature survey. *Tech. rep.*, University of Wisconsin-Madison.

Zou, H. (2006) The adaptive lasso and its oracle properties. *Journal of the American statistical association*, **101**, 1418–1429.

Zou, H., Li, R. et al. (2008) One-step sparse estimates in nonconcave penalized likelihood models. *The Annals of Statistics*, **36**, 1509–1533.

| | | $\rho = 0.2$ | | | | $\rho = 0.4$ | | | |
|---|---|---|---|---|---|---|---|---|---|
| $n$ | | $p = 10$ | | $p = 20$ | | $p = 10$ | | $p = 20$ | |
| | | $\text{APP}_{\text{SL}}$ | $\text{CV}_{\text{SL}}$ | $\text{APP}_{\text{SL}}$ | $\text{CV}_{\text{SL}}$ | $\text{APP}_{\text{SL}}$ | $\text{CV}_{\text{SL}}$ | $\text{APP}_{\text{SL}}$ | $\text{CV}_{\text{SL}}$ |
| 200 | TPR | $7.91_{8.19}$ | $0.61_{-0.77}$ | $16.30_{16.48}$ | $0.18_{-1.32}$ | $7.14_{7.64}$ | $0.43_{-0.56}$ | $15.63_{15.78}$ | $0.06_{-1.36}$ |
| | PPV | $0.62_{0.53}$ | $-0.11_{-0.36}$ | $1.38_{1.35}$ | $-0.15_{-0.33}$ | $0.54_{0.47}$ | $-0.08_{-0.29}$ | $1.25_{1.20}$ | $-0.11_{-0.29}$ |
| | NPV | $3.43_{3.61}$ | $0.45_{-0.06}$ | $6.80_{6.84}$ | $0.24_{-0.45}$ | $3.36_{3.64}$ | $0.33_{-0.07}$ | $7.19_{7.25}$ | $0.16_{-0.51}$ |
| | cut | $-3.10_{-2.51}$ | $0.40_{0.52}$ | $-6.29_{-5.79}$ | $0.55_{0.80}$ | $-3.09_{-2.52}$ | $0.42_{0.58}$ | $-6.65_{-6.05}$ | $0.56_{0.97}$ |
| | AUC | $1.32_{1.38}$ | $0.07_{-0.23}$ | $2.61_{2.63}$ | $-0.08_{-0.44}$ | $1.20_{1.28}$ | $0.05_{-0.19}$ | $2.46_{2.49}$ | $-0.06_{-0.40}$ |
| 400 | TPR | $3.39_{3.75}$ | $0.67_{-0.18}$ | $6.05_{6.31}$ | $0.43_{-0.51}$ | $3.10_{3.49}$ | $0.53_{-0.17}$ | $5.52_{5.71}$ | $0.03_{-0.76}$ |
| | PPV | $0.27_{0.25}$ | $0.01_{-0.14}$ | $0.52_{0.51}$ | $-0.02_{-0.15}$ | $0.23_{0.20}$ | $-0.01_{-0.13}$ | $0.43_{0.42}$ | $-0.05_{-0.17}$ |
| | NPV | $1.48_{1.66}$ | $0.34_{0.01}$ | $2.58_{2.67}$ | $0.24_{-0.17}$ | $1.50_{1.71}$ | $0.33_{0.03}$ | $2.64_{2.72}$ | $0.12_{-0.25}$ |
| | cut | $-1.38_{-1.17}$ | $0.03_{0.16}$ | $-2.44_{-2.25}$ | $0.15_{0.26}$ | $-1.38_{-1.16}$ | $0.08_{0.20}$ | $-2.44_{-2.19}$ | $0.28_{0.42}$ |
| | AUC | $0.59_{0.63}$ | $0.12_{-0.08}$ | $1.03_{1.07}$ | $0.07_{-0.15}$ | $0.53_{0.59}$ | $0.10_{-0.06}$ | $0.92_{0.95}$ | $0.01_{-0.17}$ |

Table 1: (**Correct Model Specification**): Percent bias of the apparent (APP) estimators and the 10 fold cross-validated (CV) estimators in the supervised (SL; subscript) and SS settings.

| $n$ | | $\rho = 0.2$ | | | | $\rho = 0.4$ | | | |
|---|---|---|---|---|---|---|---|---|---|
| | | $p = 10$ | | $p = 20$ | | $p = 10$ | | $p = 20$ | |
| | | $\text{ESE}_{\text{ASE}}$ | Cov.P | $\text{ESE}_{\text{ASE}}$ | Cov.P | $\text{ESE}_{\text{ASE}}$ | Cov.P | $\text{ESE}_{\text{ASE}}$ | Cov.P |
| 200 | TPR | $6.60_{7.20}$ | 0.945 | $6.75_{7.94}$ | 0.957 | $6.10_{6.72}$ | 0.944 | $6.31_{7.66}$ | 0.969 |
| | PPV | $1.22_{1.12}$ | 0.930 | $1.35_{1.16}$ | 0.919 | $1.04_{0.97}$ | 0.928 | $1.09_{1.03}$ | 0.932 |
| | NPV | $3.28_{3.79}$ | 0.963 | $3.27_{4.17}$ | 0.978 | $3.20_{3.73}$ | 0.965 | $3.26_{4.26}$ | 0.980 |
| | cut | $2.94_{3.09}$ | 0.947 | $3.02_{3.41}$ | 0.957 | $2.86_{3.03}$ | 0.946 | $2.85_{3.42}$ | 0.965 |
| | AUC | $1.33_{1.42}$ | 0.933 | $1.46_{1.57}$ | 0.955 | $1.20_{1.26}$ | 0.943 | $1.29_{1.45}$ | 0.968 |
| 400 | TPR | $4.74_{4.87}$ | 0.953 | $4.77_{5.02}$ | 0.948 | $4.45_{4.51}$ | 0.939 | $4.51_{4.69}$ | 0.947 |
| | PPV | $0.83_{0.78}$ | 0.929 | $0.81_{0.78}$ | 0.939 | $0.71_{0.68}$ | 0.919 | $0.74_{0.68}$ | 0.924 |
| | NPV | $2.39_{2.53}$ | 0.959 | $2.41_{2.61}$ | 0.953 | $2.36_{2.48}$ | 0.954 | $2.36_{2.58}$ | 0.963 |
| | cut | $2.14_{2.12}$ | 0.945 | $2.08_{2.16}$ | 0.949 | $2.07_{2.02}$ | 0.931 | $2.11_{2.10}$ | 0.935 |
| | AUC | $0.91_{0.92}$ | 0.930 | $0.94_{0.95}$ | 0.939 | $0.81_{0.82}$ | 0.937 | $0.84_{0.84}$ | 0.935 |

Table 2: (**Correct Model Specification**): Empirical standard error (ESE), median of the estimated standard errors using perturbation resampling (ASE, subscript), and the coverage probabilities of the 95% confidence intervals using the ASE. The ESE and ASE are multiplied by 100.

(a) Algorithm Accuracy for EMR-based RA Study

| | Semi-Supervised | | | Supervised | | | |
|---|---|---|---|---|---|---|---|
| | APP | CV | ASE | APP | CV | ASE | RE |
| TPR | 86.70 | 73.86 | 7.73 | 87.90 | 74.50 | 10.59 | **1.88** |
| PPV | 79.47 | 76.79 | 2.43 | 81.28 | 78.63 | 3.89 | **2.56** |
| NPV | 96.97 | 94.19 | 1.78 | 96.95 | 93.79 | 2.42 | **1.85** |
| Cut | 80.09 | 82.62 | 1.64 | 78.79 | 80.95 | 3.05 | **3.45** |
| AUC | 97.65 | 94.93 | 2.82 | 97.77 | 94.51 | 4.86 | **2.97** |

(b) Algorithm Accuracy for EMR-based MS Study

| | Semi-Supervised | | | Supervised | | | |
|---|---|---|---|---|---|---|---|
| | APP | CV | ASE | APP | CV | ASE | RE |
| TPR | 79.49 | 76.93 | 4.20 | 77.58 | 73.39 | 6.17 | **2.16** |
| PPV | 94.21 | 94.03 | 0.41 | 94.36 | 94.05 | 0.73 | **3.23** |
| NPV | 81.91 | 80.10 | 3.25 | 79.74 | 76.82 | 4.38 | **1.81** |
| Cut | 57.34 | 58.80 | 2.18 | 57.57 | 59.47 | 3.54 | **2.64** |
| AUC | 94.19 | 93.94 | 0.99 | 93.58 | 92.88 | 1.26 | **1.61** |

Table 3: Apparent (APP) and 10-fold cross-validated (CV) estimates of the SS and supervised accuracy measures along with their estimated standard errors (ASE) and relative efficiencies (RE; Supervised:Semi-Supervised) for the EMR-based studies of RA and MS. All values are multiplied by 100.

## 1A: **Correct Specification**

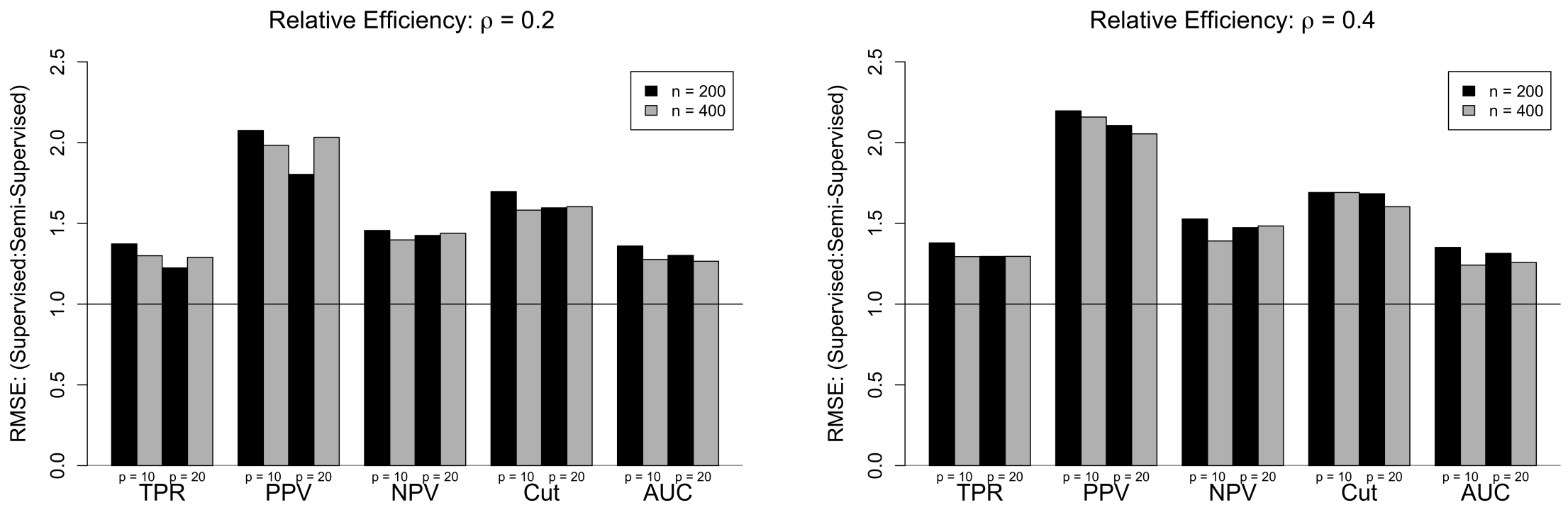

## 1B: **Misspecified Link Function**

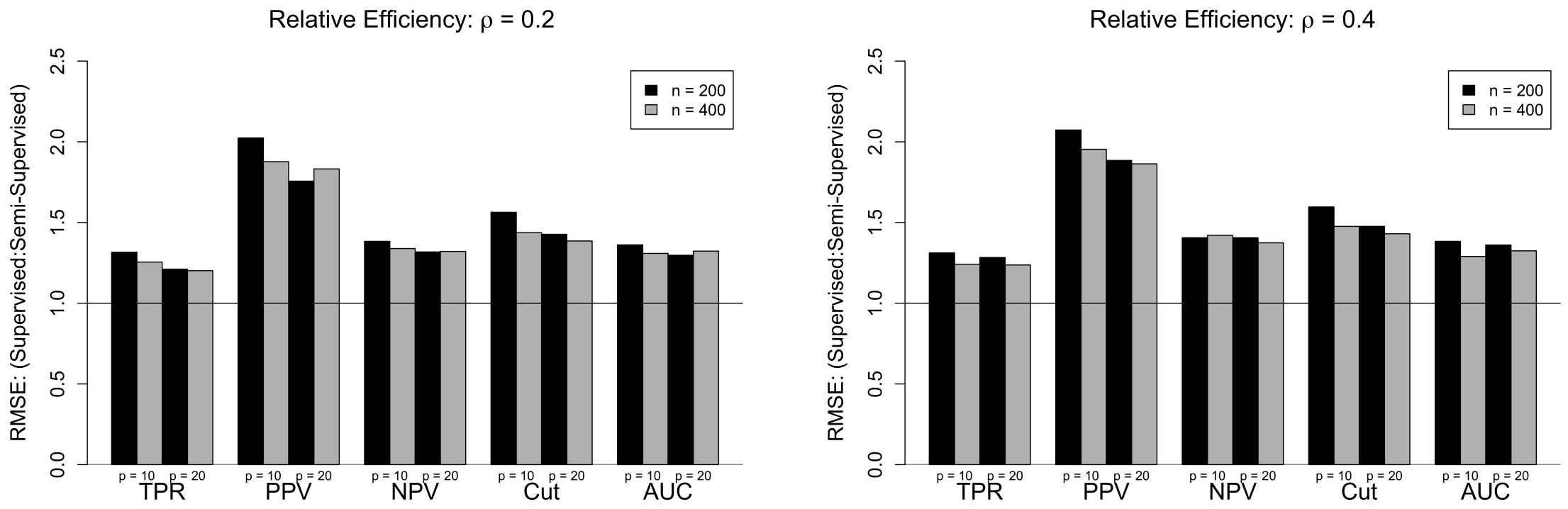

## 1C: **Misspecified Linear Predictor**

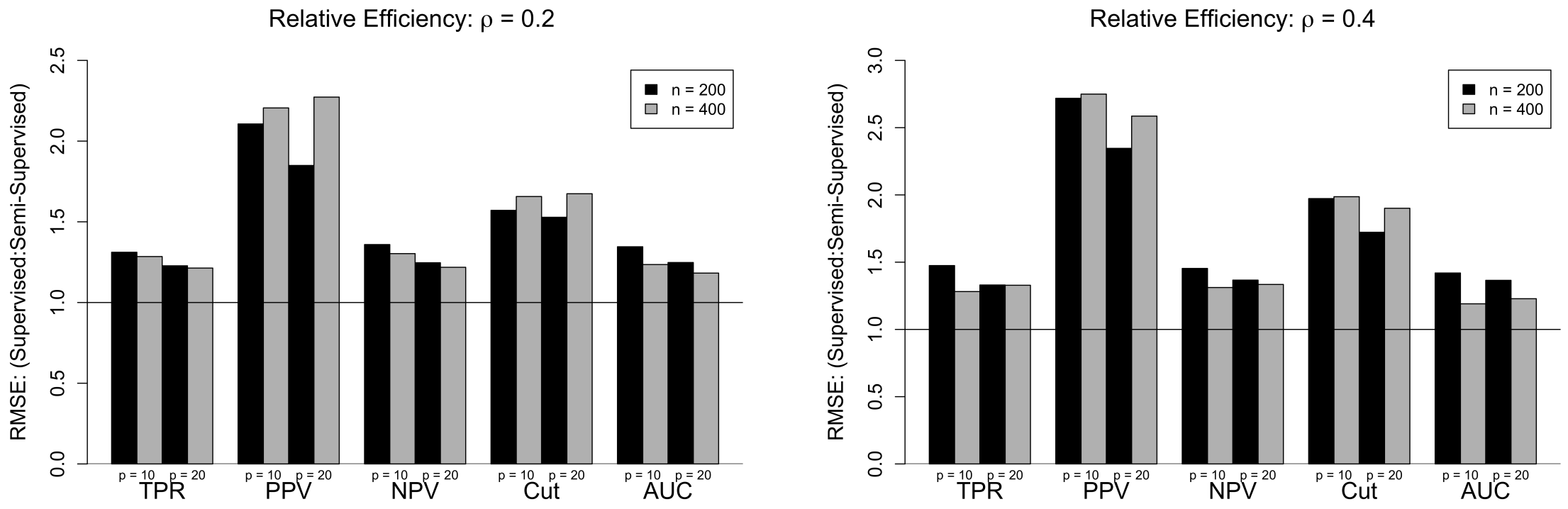

Figure 1: Efficiency of the 10 fold cross-validated SS estimators relative to their supervised counterparts with respect to mean squared error.

# Supplementary Materials for 'Semi-Supervised Approaches to Efficient Evaluation of Model Prediction Performance'

Jessica Gronsbell & Tianxi Cai

# 1. Additional Simulation Results

A: **Misspecified Link Function**

| | | $\rho = 0.2$ | | | | $\rho = 0.4$ | | | |
|---|---|---|---|---|---|---|---|---|---|
| $n$ | | $p = 10$ | | $p = 20$ | | $p = 10$ | | $p = 20$ | |
| | | $APP_{SL}$ | $CV_{SL}$ | $APP_{SL}$ | $CV_{SL}$ | $APP_{SL}$ | $CV_{SL}$ | $APP_{SL}$ | $CV_{SL}$ |
| 200 | TPR | $7.57_{7.47}$ | $1.16_{-0.72}$ | $15.85_{15.54}$ | $1.26_{-0.62}$ | $7.11_{6.78}$ | $0.85_{-0.98}$ | $14.29_{13.81}$ | $0.39_{-1.58}$ |
| | PPV | $0.34_{0.27}$ | $0.02_{-0.18}$ | $0.70_{0.65}$ | $-0.02_{-0.18}$ | $0.31_{0.27}$ | $0.01_{-0.14}$ | $0.66_{0.62}$ | $-0.01_{-0.16}$ |
| | NPV | $5.42_{5.59}$ | $0.94_{0.02}$ | $11.37_{11.26}$ | $1.18_{0.10}$ | $5.52_{5.31}$ | $0.86_{-0.33}$ | $10.81_{10.42}$ | $0.44_{-0.87}$ |
| | cut | $-4.91_{-3.94}$ | $-0.21_{0.73}$ | $-9.62_{-8.66}$ | $-0.02_{0.94}$ | $-4.89_{-3.94}$ | $-0.04_{0.86}$ | $-9.55_{-8.57}$ | $0.20_{1.33}$ |
| | AUC | $1.50_{1.51}$ | $0.15_{-0.30}$ | $2.96_{2.93}$ | $0.13_{-0.36}$ | $1.40_{1.37}$ | $0.13_{-0.29}$ | $2.67_{2.64}$ | $-0.01_{-0.46}$ |
| 400 | TPR | $3.36_{3.36}$ | $0.92_{-0.24}$ | $5.69_{5.63}$ | $0.51_{-0.62}$ | $3.28_{3.51}$ | $0.82_{0.02}$ | $5.33_{5.38}$ | $0.18_{-0.73}$ |
| | PPV | $0.15_{0.11}$ | $0.02_{-0.09}$ | $0.27_{0.25}$ | $0.00_{-0.08}$ | $0.14_{0.14}$ | $0.02_{-0.04}$ | $0.23_{0.22}$ | $-0.03_{-0.10}$ |
| | NPV | $2.47_{2.62}$ | $0.81_{0.17}$ | $4.00_{3.97}$ | $0.43_{-0.30}$ | $2.52_{2.68}$ | $0.72_{0.12}$ | $4.11_{4.17}$ | $0.35_{-0.29}$ |
| | cut | $-2.20_{-1.70}$ | $-0.28_{0.38}$ | $-3.79_{-3.39}$ | $-0.12_{0.46}$ | $-2.29_{-2.09}$ | $-0.27_{0.09}$ | $-3.63_{-3.27}$ | $0.27_{0.78}$ |
| | AUC | $0.66_{0.67}$ | $0.15_{-0.14}$ | $1.10_{1.10}$ | $0.04_{-0.24}$ | $0.61_{0.66}$ | $0.12_{-0.08}$ | $1.02_{1.03}$ | $0.00_{-0.23}$ |

**B: Misspecified Linear Predictor**

| | | $APP_{SL}$ | $CV_{SL}$ | $APP_{SL}$ | $CV_{SL}$ | $APP_{SL}$ | $CV_{SL}$ | $APP_{SL}$ | $CV_{SL}$ |
|---|---|---|---|---|---|---|---|---|---|
| 200 | TPR | $5.99_{6.61}$ | $-0.34_{-1.71}$ | $12.98_{13.76}$ | $-0.01_{-1.69}$ | $4.76_{5.42}$ | $-0.27_{-1.44}$ | $10.59_{11.60}$ | $-0.58_{-1.97}$ |
| | PPV | $0.42_{0.34}$ | $-0.11_{-0.31}$ | $0.87_{0.88}$ | $-0.15_{-0.28}$ | $0.29_{0.26}$ | $-0.10_{-0.21}$ | $0.65_{0.66}$ | $-0.16_{-0.25}$ |
| | NPV | $3.26_{3.81}$ | $-0.01_{-0.64}$ | $6.95_{7.46}$ | $0.26_{-0.71}$ | $3.13_{3.67}$ | $0.08_{-0.66}$ | $6.63_{7.35}$ | $-0.06_{-1.01}$ |
| | cut | $-3.38_{-2.72}$ | $0.73_{1.02}$ | $-6.48_{-6.20}$ | $0.82_{0.89}$ | $-2.99_{-2.61}$ | $0.89_{1.01}$ | $-6.23_{-6.13}$ | $1.14_{1.18}$ |
| | AUC | $1.21_{1.23}$ | $0.19_{-0.43}$ | $2.45_{2.51}$ | $0.21_{-0.42}$ | $1.01_{1.05}$ | $0.24_{-0.31}$ | $1.97_{2.04}$ | $0.12_{-0.47}$ |
| 400 | TPR | $2.56_{2.86}$ | $0.24_{-0.75}$ | $4.69_{5.24}$ | $0.25_{-0.68}$ | $2.21_{2.60}$ | $0.23_{-0.40}$ | $3.81_{4.40}$ | $0.09_{-0.56}$ |
| | PPV | $0.15_{0.11}$ | $-0.05_{-0.16}$ | $0.34_{0.32}$ | $-0.02_{-0.12}$ | $0.14_{0.14}$ | $-0.01_{-0.06}$ | $0.24_{0.25}$ | $-0.04_{-0.08}$ |
| | NPV | $1.49_{1.73}$ | $0.28_{-0.24}$ | $2.61_{2.99}$ | $0.25_{-0.25}$ | $1.47_{1.74}$ | $0.24_{-0.19}$ | $2.46_{2.88}$ | $0.14_{-0.30}$ |
| | cut | $-1.38_{-1.09}$ | $0.28_{0.55}$ | $-2.62_{-2.48}$ | $0.17_{0.31}$ | $-1.47_{-1.35}$ | $0.15_{0.24}$ | $-2.41_{-2.41}$ | $0.32_{0.32}$ |
| | AUC | $0.50_{0.48}$ | $0.15_{-0.24}$ | $0.96_{0.98}$ | $0.23_{-0.16}$ | $0.54_{0.52}$ | $0.28_{-0.06}$ | $0.79_{0.79}$ | $0.23_{-0.12}$ |

Table 1S: Percent bias of the apparent (APP) estimators and the 10 fold cross-validated (CV) estimators in the supervised (SL; subscript) and SS settings.

A: **Misspecified Link Function**

| | | $\rho = 0.2$ | | | | $\rho = 0.4$ | | | |
|---|---|---|---|---|---|---|---|---|---|
| $n$ | | $p = 10$ | | $p = 20$ | | $p = 10$ | | $p = 20$ | |
| | | $\text{ESE}_{\text{ASE}}$ | Cov.P | $\text{ESE}_{\text{ASE}}$ | Cov.P | $\text{ESE}_{\text{ASE}}$ | Cov.P | $\text{ESE}_{\text{ASE}}$ | Cov.P |
| 200 | TPR | $6.65_{7.15}$ | 0.946 | $6.83_{7.71}$ | 0.955 | $6.09_{6.69}$ | 0.945 | $6.38_{7.44}$ | 0.964 |
| | PPV | $0.72_{0.67}$ | 0.918 | $0.78_{0.68}$ | 0.921 | $0.64_{0.60}$ | 0.921 | $0.71_{0.63}$ | 0.917 |
| | NPV | $4.88_{5.63}$ | 0.964 | $4.94_{6.11}$ | 0.973 | $4.63_{5.46}$ | 0.973 | $4.63_{6.07}$ | 0.986 |
| | cut | $3.97_{4.21}$ | 0.947 | $4.05_{4.51}$ | 0.959 | $3.74_{3.97}$ | 0.948 | $3.91_{4.37}$ | 0.959 |
| | AUC | $1.51_{1.58}$ | 0.934 | $1.58_{1.71}$ | 0.949 | $1.30_{1.42}$ | 0.944 | $1.41_{1.55}$ | 0.952 |
| 400 | TPR | $4.72_{4.84}$ | 0.940 | $4.93_{4.96}$ | 0.933 | $4.18_{4.52}$ | 0.944 | $4.54_{4.66}$ | 0.947 |
| | PPV | $0.48_{0.47}$ | 0.935 | $0.50_{0.47}$ | 0.926 | $0.43_{0.42}$ | 0.940 | $0.45_{0.42}$ | 0.933 |
| | NPV | $3.49_{3.71}$ | 0.951 | $3.54_{3.82}$ | 0.959 | $3.26_{3.66}$ | 0.963 | $3.46_{3.76}$ | 0.963 |
| | cut | $2.82_{2.84}$ | 0.940 | $2.95_{2.93}$ | 0.925 | $2.56_{2.69}$ | 0.945 | $2.73_{2.75}$ | 0.941 |
| | AUC | $1.00_{1.04}$ | 0.929 | $1.05_{1.05}$ | 0.938 | $0.85_{0.92}$ | 0.951 | $0.95_{0.95}$ | 0.941 |

B: **Misspecified Linear Predictor**

| | | $\text{ESE}_{\text{ASE}}$ | Cov.P | $\text{ESE}_{\text{ASE}}$ | Cov.P | $\text{ESE}_{\text{ASE}}$ | Cov.P | $\text{ESE}_{\text{ASE}}$ | Cov.P |
|---|---|---|---|---|---|---|---|---|---|
| 200 | TPR | $6.90_{7.72}$ | 0.957 | $7.26_{8.26}$ | 0.961 | $5.75_{6.49}$ | 0.957 | $6.01_{7.26}$ | 0.968 |
| | PPV | $0.94_{0.87}$ | 0.939 | $1.04_{0.90}$ | 0.931 | $0.67_{0.67}$ | 0.949 | $0.75_{0.73}$ | 0.948 |
| | NPV | $4.19_{5.02}$ | 0.969 | $4.32_{5.37}$ | 0.979 | $3.96_{4.72}$ | 0.957 | $4.02_{5.23}$ | 0.977 |
| | cut | $3.41_{3.67}$ | 0.955 | $3.47_{3.88}$ | 0.967 | $2.92_{3.23}$ | 0.959 | $3.00_{3.62}$ | 0.977 |
| | AUC | $1.84_{2.01}$ | 0.912 | $2.02_{2.08}$ | 0.916 | $1.62_{1.77}$ | 0.903 | $1.70_{1.88}$ | 0.939 |
| 400 | TPR | $4.53_{5.02}$ | 0.950 | $4.77_{5.12}$ | 0.944 | $3.75_{4.19}$ | 0.959 | $4.01_{4.33}$ | 0.955 |
| | PPV | $0.58_{0.58}$ | 0.947 | $0.60_{0.58}$ | 0.936 | $0.43_{0.43}$ | 0.943 | $0.45_{0.44}$ | 0.940 |
| | NPV | $2.87_{3.26}$ | 0.958 | $2.99_{3.34}$ | 0.963 | $2.69_{3.04}$ | 0.968 | $2.83_{3.16}$ | 0.960 |
| | cut | $2.24_{2.38}$ | 0.953 | $2.32_{2.42}$ | 0.944 | $1.93_{2.06}$ | 0.953 | $2.03_{2.14}$ | 0.952 |
| | AUC | $1.19_{1.32}$ | 0.949 | $1.30_{1.33}$ | 0.924 | $1.07_{1.16}$ | 0.911 | $1.11_{1.17}$ | 0.927 |

Table 2S: Empirical standard error (ESE), median of the estimated standard errors using perturbation resampling (ASE, subscript), and the coverage probabilities of the 95% confidence intervals using the ASE. The ESE and ASE are multiplied by 100.

| | **Semi-supervised** | | | | **Supervised** | | | |
|---|---|---|---|---|---|---|---|---|
| | ALASSO | | GLM | | ALASSO | | GLM | |
| | $\text{ESE}_{\text{ASE}}$ | Cov.P | $\text{ESE}_{\text{ASE}}$ | Cov.P | $\text{ESE}_{\text{ASE}}$ | Cov.P | $\text{ESE}_{\text{ASE}}$ | Cov.P |
| TPR | $6.31_{7.66}$ | 0.969 | $6.56_{11.25}$ | 0.996 | $7.11_{9.90}$ | 0.988 | $7.04_{11.49}$ | 0.993 |
| PPV | $1.09_{1.03}$ | 0.932 | $1.26_{1.47}$ | 0.981 | $1.57_{1.92}$ | 0.983 | $1.72_{2.33}$ | 0.991 |
| NPV | $3.26_{4.26}$ | 0.980 | $3.25_{5.74}$ | 0.998 | $3.94_{5.05}$ | 0.985 | $3.86_{5.50}$ | 0.990 |
| cut | $2.85_{3.42}$ | 0.965 | $2.92_{4.67}$ | 0.994 | $3.68_{4.47}$ | 0.974 | $3.65_{5.19}$ | 0.981 |
| AUC | $1.29_{1.45}$ | 0.968 | $1.38_{2.39}$ | 0.999 | $1.43_{2.04}$ | 0.989 | $1.46_{2.71}$ | 0.999 |

Table 3S: Empirical standard error (ESE), median of the estimated standard errors using perturbation resampling (ASE, subscript), and the coverage probabilities of the 95% confidence intervals using the ASE with a GLM without penalization versus ALASSO when $p = 20$, $\rho = 0.4$, and $n = 200$ under correct model specification in both the supervised and SS settings. The ESE and ASE are multiplied by 100. (Results for the GLM without penalization based on the 1341/1500 simulations without convergence issues.)

| | $\overline{\text{TPR}}$ | $\overline{\text{PPV}}$ | $\overline{\text{NPV}}$ | $\overline{\text{AUC}}$ |
|---|---|---|---|---|
| GLM w/o penalization | 69.84 | 90.02 | 82.98 | 94.07 |
| ALASSO | 73.34 | 90.45 | 84.64 | 94.91 |

Table 4S: Target parameters based on the estimator from a GLM without penalization vs ALASSO when $p = 20$, $\rho = 0.4$, and $n = 200$ under correct model specification. All values multiplied by 100.

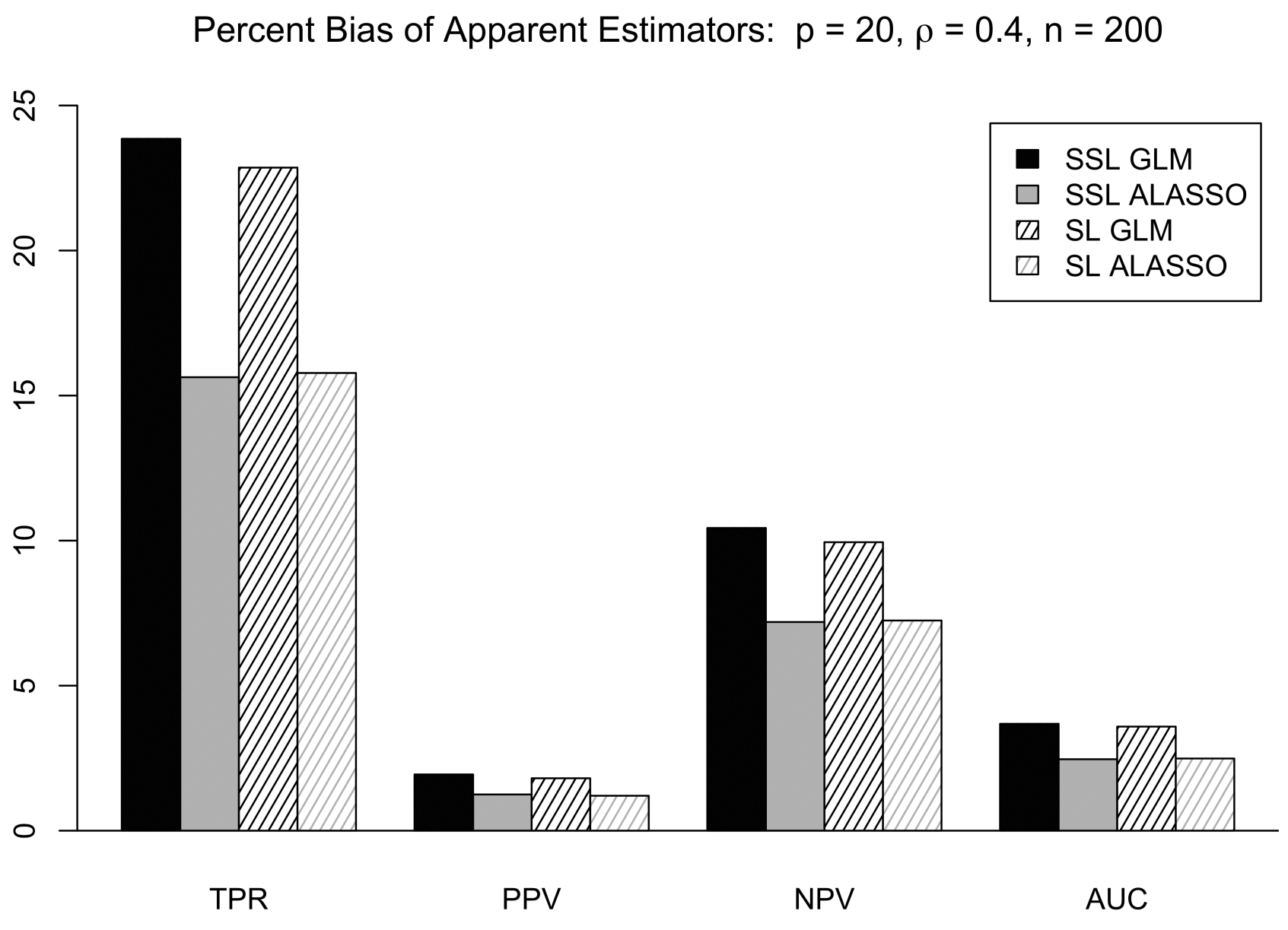

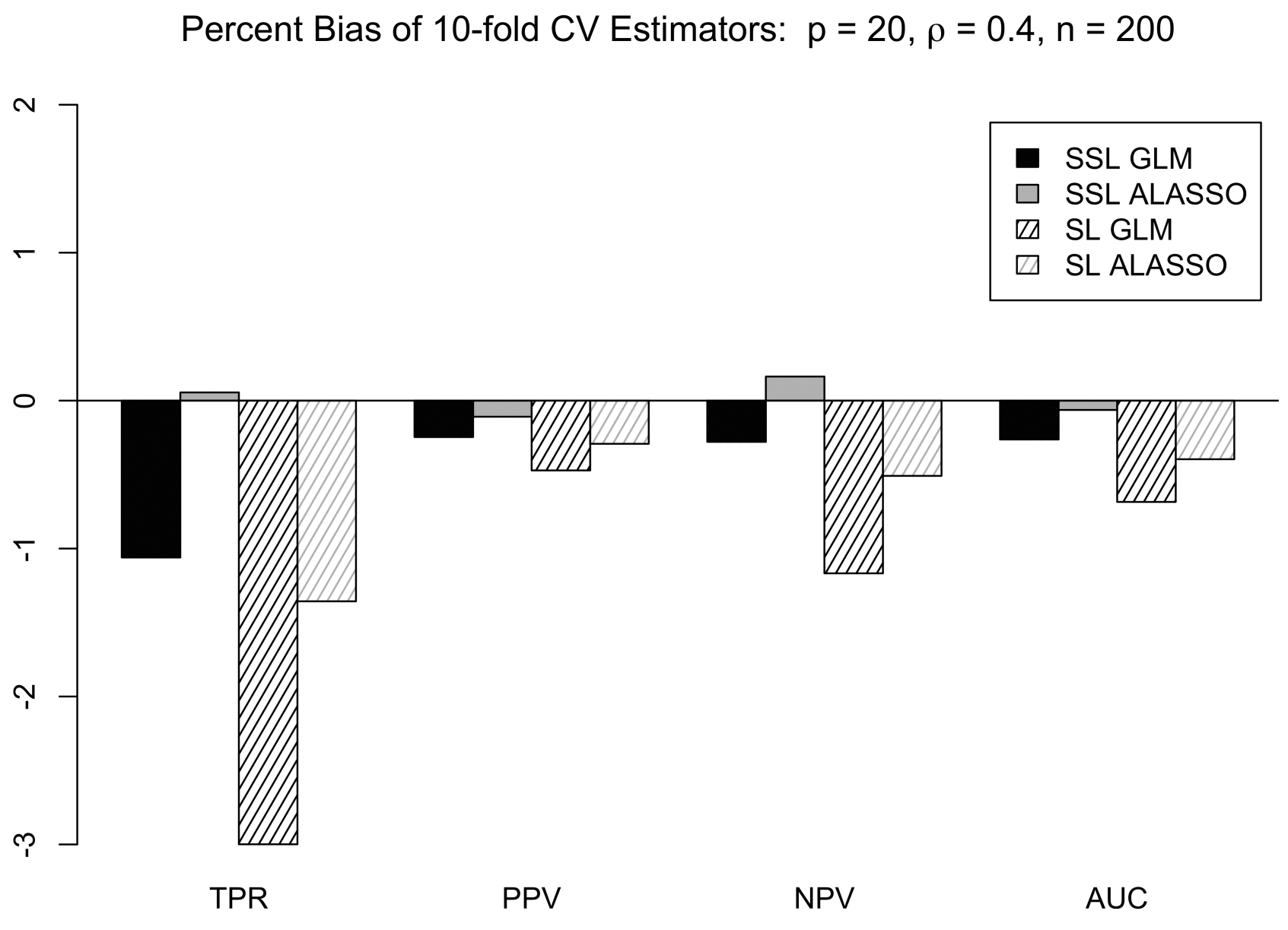

Figure 1S: Percent bias of the supervised and SS estimators using a GLM without penalization versus ALASSO with and without CV when $p = 20$, $\rho = 0.4$, and $n = 200$ under correct model specification. (Results for the GLM without penalization based on the 1341/1500 simulations without convergence issues.)

# Appendix

Throughout, we assume that $\Omega$ is a compact parameter space containing $\boldsymbol{\theta}_0$ and that the covariates $\mathbf{x}$ are bounded. Without loss of generality, we further assume that $\boldsymbol{\theta}_0^{\mathsf{T}}\mathbf{x}$ is a continuous random variable with continuously differentiable density function. We also require that $P(Y = 1 \mid \mathbf{x})$ is twice continuously differentiable and denote by $\mathbb{P}^*$ the measure generated by $\mathcal{D}$ and the perturbation variables $\mathcal{G}$.

In addition to the expansion given in (2.3), $E(\widehat{\boldsymbol{\theta}} - \boldsymbol{\theta}_0) = o(n^{-\frac{1}{2}})$. It then follows that

$$\sup_c \left\{ |\bar{m}(c) - m(c, \mathcal{P}_{\boldsymbol{\theta}_0})| + \left|\overline{\mathrm{TPR}}(c) - \mathrm{TPR}(c, \mathcal{P}_{\boldsymbol{\theta}_0})\right| + \left|\overline{\mathrm{FPR}}(c) - \mathrm{FPR}(c, \mathcal{P}_{\boldsymbol{\theta}_0})\right| \right\} = o(n^{-\frac{1}{2}}),$$

where $\mathcal{P}_{\boldsymbol{\theta}}(\mathbf{x}) = F_{\boldsymbol{\theta}}(\boldsymbol{\theta}^{\mathsf{T}}\mathbf{x})$, $m(c, \mathcal{P}_{\boldsymbol{\theta}}) = P\{y = 1 \mid \mathcal{P}_{\boldsymbol{\theta}}(\mathbf{x}) = c\}$, $\mathrm{TPR}(c, \mathcal{P}_{\boldsymbol{\theta}}) = P\{\mathcal{P}_{\boldsymbol{\theta}}(\mathbf{x}) \geq c \mid y = 1\}$ and $\mathrm{FPR}(c, \mathcal{P}_{\boldsymbol{\theta}}) = P\{\mathcal{P}_{\boldsymbol{\theta}}(\mathbf{x}) \geq c \mid y = 0\}$. Here we note that $m(c, \mathcal{P}_{\boldsymbol{\theta}_0}) = m(s)$ and $\mathcal{P}_{\boldsymbol{\theta}}(\mathbf{x})$ follows a uniform distribution. Additionally, we assume that $h = O(n^{-\nu})$, $\nu \in (1/4, 1/2)$ and $h^{-1} = O(N^{\kappa})$ for $\kappa < 1/4$.

## A. Asymptotic Properties of $\widehat{\mathbf{TPR}}(c)$ and $\widehat{\mathbf{ROC}}(u_0)$

To establish the uniform consistency of $\widehat{\mathrm{TPR}}(c)$, we write $\widehat{\mathrm{TPR}}(c) = \widehat{\xi}(c, \widehat{\boldsymbol{\theta}})/\widehat{\xi}(0, \widehat{\boldsymbol{\theta}})$, $\mathrm{TPR}(c) = \xi(c, \boldsymbol{\theta}_0)/\xi(0, \boldsymbol{\theta}_0)$, where

$$\widehat{\xi}(c, \boldsymbol{\theta}) = N^{-1} \sum_{i=1}^{N} I\{\widehat{\mathcal{P}}_{\boldsymbol{\theta}}(\mathbf{x}_i) \geq c\} \widetilde{m}\{\widehat{\mathcal{P}}_{\boldsymbol{\theta}}(\mathbf{x}_i), \widehat{\mathcal{P}}_{\boldsymbol{\theta}}\} = \int_c^1 \widetilde{m}(s, \widehat{\mathcal{P}}_{\boldsymbol{\theta}}) \widehat{\mathcal{V}}(ds, \widehat{\mathcal{P}}_{\boldsymbol{\theta}}),$$

$$\xi(c, \boldsymbol{\theta}) = E[I\{\mathcal{P}_{\boldsymbol{\theta}}(\mathbf{x}) \geq c\} m\{\mathcal{P}_{\boldsymbol{\theta}}(\mathbf{x}), \mathcal{P}_{\boldsymbol{\theta}}\}] = \int_c^1 m(s, \mathcal{P}_{\boldsymbol{\theta}}) \mathcal{V}(ds, \mathcal{P}_{\boldsymbol{\theta}})$$

$\widehat{\mathcal{V}}(s, \widehat{\mathcal{P}}_{\boldsymbol{\theta}}) = N^{-1} \sum_{i=1}^{N} I\{\widehat{\mathcal{P}}_{\boldsymbol{\theta}}(\mathbf{x}_i) \leq s\}$, $\mathcal{V}(s, \mathcal{P}_{\boldsymbol{\theta}}) = P\{\mathcal{P}_{\boldsymbol{\theta}}(\mathbf{x}) \leq s\} = s$, $\widehat{\mathcal{P}}_{\boldsymbol{\theta}}(\mathbf{x}) = \widehat{F}_{\boldsymbol{\theta}}(\boldsymbol{\theta}^{\mathsf{T}}\mathbf{x})$, and $\widetilde{m}(s, \widehat{\mathcal{P}}_{\boldsymbol{\theta}}) = [\sum_{i=1}^{n} K_h\{\widehat{\mathcal{P}}_{\boldsymbol{\theta}}(\mathbf{x}_i) - s\} y_i]/[\sum_{i=1}^{n} K_h\{\widehat{\mathcal{P}}_{\boldsymbol{\theta}}(\mathbf{x}_i) - s\}]$. Thus, it suffices to show that

$$\sup_s |\mathcal{V}(s, \mathcal{P}_{\widehat{\boldsymbol{\theta}}}) - \mathcal{V}(s, \mathcal{P}_{\boldsymbol{\theta}_0})| + \sup_s |\widetilde{m}(s, \widehat{\mathcal{P}}_{\widehat{\boldsymbol{\theta}}}) - m(c, \mathcal{P}_{\boldsymbol{\theta}_0})| \to 0, \quad \text{in probability.}$$

From standard empirical process theory (Pollard, 1990), one may show that $\sup_{\boldsymbol{\theta}, \mathbf{x}} |\widehat{\mathcal{P}}_{\boldsymbol{\theta}}(\mathbf{x}) - \mathcal{P}_{\boldsymbol{\theta}}(\mathbf{x})| = O_p(N^{-1/2})$ and $N^{1/2}\{\widehat{\mathcal{V}}(s, \widehat{\mathcal{P}}_{\boldsymbol{\theta}}) - \mathcal{V}(s, \mathcal{P}_{\boldsymbol{\theta}})\}$ converges weakly to a zero-mean Gaussian process in $(s, \boldsymbol{\theta})$. This, together with the consistency of $\widehat{\boldsymbol{\theta}}$, implies that $\sup_s |\mathcal{V}(s, \mathcal{P}_{\widehat{\boldsymbol{\theta}}}) -$

$\mathcal{V}(s, \mathcal{P}_{\boldsymbol{\theta}_0})| \to 0$ in probability. For the consistency of $\widetilde{m}(s, \widehat{\mathcal{P}}_{\widehat{\boldsymbol{\theta}}})$, we write

$$n^{-1}\sum_{i=1}^{n} K_h\{\widehat{\mathcal{P}}_{\widehat{\boldsymbol{\theta}}}(\mathbf{x}_i) - s\}\{y_i - m(s)\} = \int K_h(t-s)\ \widehat{\mathcal{H}}(dt, s)$$

where $\widehat{\mathcal{H}}(t,s) = n^{-1}\sum_{i=1}^{n} I\{\widehat{\mathcal{P}}_{\widehat{\boldsymbol{\theta}}}(\mathbf{x}_i) \le t\}\{y_i - m(s)\}$. Similarly, it is not difficult to show that $n^{\frac{1}{2}}\{\widehat{\mathcal{H}}(t,s) - \mathcal{H}_0(t,s)\}$ converges weakly to a zero-mean Gaussian process, where $\mathcal{H}_0(t,s) = E[I\{\mathcal{P}_{\boldsymbol{\theta}_0}(\mathbf{x}_i) \le t\}\{y_i - m(s)\}]$. It follows that uniformly in $s$,

$$n^{-1}\sum_{i=1}^{n} K_h\{\widehat{\mathcal{P}}_{\widehat{\boldsymbol{\theta}}}(\mathbf{x}_i) - s\}\{y_i - m(s)\} = \int K_h(t-s)\mathcal{H}_0(dt, s) + O_p(n^{-\frac{1}{2}}h^{-1}) \to 0, \text{ in probability.}$$

Similar arguments can be used to show that $\sup_s |n^{-1}\sum_{i=1}^{n} K_h\{\widehat{\mathcal{P}}_{\widehat{\boldsymbol{\theta}}}(\mathbf{x}_i) - s\} - 1| \to 0$ uniformly in $x$, where the constant 1 is the density of $\mathcal{V}(s, \mathcal{P}_{\boldsymbol{\theta}_0}) = s$. Therefore $\sup_s |\widetilde{m}(s, \widehat{\mathcal{P}}_{\widehat{\boldsymbol{\theta}}}) - m(c, \mathcal{P}_{\boldsymbol{\theta}_0})| \to 0$ in probability and hence we conclude the uniform consistency of $\widehat{\mathrm{TPR}}(c)$.

For the weak convergence of $\widehat{\mathcal{W}}_{\mathrm{TPR}}(c) = n^{\frac{1}{2}}\{\widehat{\mathrm{TPR}}(c) - \mathrm{TPR}(c)\}$, it suffices to derive the asymptotic expansions for $\widehat{\mathcal{W}}_{\xi}(\boldsymbol{\theta}, c) = n^{\frac{1}{2}}\{\widehat{\xi}(\boldsymbol{\theta}, c) - \xi(\boldsymbol{\theta}, c)\}$. To this end, we first note that

$$\begin{aligned}
& n^{\frac{1}{2}} \sup_{s,\boldsymbol{\theta}} \left| n^{-1}\sum_{i=1}^{n} K_h\{\widehat{\mathcal{P}}_{\boldsymbol{\theta}}(\mathbf{x}_i) - s\} y_i^a - n^{-1}\sum_{i=1}^{n} K_h\{\mathcal{P}_{\boldsymbol{\theta}}(\mathbf{x}_i) - s\} y_i^a \right| \\
&= \sup_{s,\boldsymbol{\theta}} \left| \int K_h(v-s) d\mathbb{G}_n \left[ I\{\boldsymbol{\theta}^{\mathsf{T}}\mathbf{x} \le \widehat{F}_{\boldsymbol{\theta}}^{-1}(s)\} y^a - I\{\boldsymbol{\theta}^{\mathsf{T}}\mathbf{x} \le F_{\boldsymbol{\theta}}^{-1}(s)\} y^a \right] \right| \\
&+ n^{\frac{1}{2}} \sup_{s,\boldsymbol{\theta}} \left| \int K_h(v-s) d\mathbb{P} \left[ I\{\boldsymbol{\theta}^{\mathsf{T}}\mathbf{x} \le \widehat{F}_{\boldsymbol{\theta}}^{-1}(s)\} y^a - I\{\boldsymbol{\theta}^{\mathsf{T}}\mathbf{x} \le F_{\boldsymbol{\theta}}^{-1}(s)\} y^a \right] \right| \\
&\le h^{-1}\|\mathbb{G}_n\|_{\mathcal{H}_\delta} + O_p\{(n/N)^{\frac{1}{2}}\}
\end{aligned}$$

where $\mathcal{H}_\delta = \{I(\theta^{\mathsf{T}}\mathbf{x} \le s')y^a - I(\theta^{\mathsf{T}}\mathbf{x} \le s)y^a : \boldsymbol{\theta}, |s - s'| \le \delta\}$ is a class of functions indexed by $\boldsymbol{\theta}$, $s$, $\delta$ and $a = 0$ or 1. Furthermore, $\mathcal{H}_\delta$ is uniformly bounded by an envelop function of order $\delta^{1/2}$ with respect to $L_2$ norm. By the maximum inequality (Theorem 2.14.2, Van der Vaart & Wellner, 1996) and $\sup_{s,\boldsymbol{\theta}} |\widehat{F}_{\boldsymbol{\theta}}^{-1}(s) - F_{\boldsymbol{\theta}}^{-1}(s)| = O_p(N^{-1/2})$, we have $h^{-1}\|\mathbb{G}_n\|_{\mathcal{H}_\delta} \lesssim O_p\{h^{-1}N^{-1/4}\log(N)\} = o_p(1)$. It follows that $\sup_{s,\boldsymbol{\theta}} |n^{\frac{1}{2}}\{\widetilde{m}(s, \widehat{\mathcal{P}}_{\boldsymbol{\theta}}) - \widetilde{m}(s, \mathcal{P}_{\boldsymbol{\theta}})\}| = o_p(1)$ and

consequently

$$\widehat{\mathcal{W}}_\xi(\boldsymbol{\theta}, c) = n^{\frac{1}{2}} \int_c^1 \left\{ \widetilde{m}(s, \mathcal{P}_{\boldsymbol{\theta}}) \widehat{\mathcal{V}}(ds, \mathcal{P}_{\boldsymbol{\theta}}) - m(s, \mathcal{P}_{\boldsymbol{\theta}}) \mathcal{V}(ds, \mathcal{P}_{\boldsymbol{\theta}}) \right\} + o_p(1)$$
$$= \widehat{\mathcal{E}} + n^{\frac{1}{2}} \int_c^1 m(s, \mathcal{P}_{\boldsymbol{\theta}}) \left\{ \widehat{\mathcal{V}}(ds, \mathcal{P}_{\boldsymbol{\theta}}) - \mathcal{V}(ds, \mathcal{P}_{\boldsymbol{\theta}}) \right\} + o_p(1)$$

where $\widehat{\mathcal{E}} = n^{\frac{1}{2}} \mathbb{P}_N(I\{\mathcal{P}_{\boldsymbol{\theta}}(\mathbf{x}) \geq c\}[\widetilde{m}\{\mathcal{P}_{\boldsymbol{\theta}}(\mathbf{x}), \mathcal{P}_{\boldsymbol{\theta}}\} - m\{\mathcal{P}_{\boldsymbol{\theta}}(\mathbf{x}), \mathcal{P}_{\boldsymbol{\theta}}\}])$ and $\mathbb{P}_N$ is the empirical measure generated by $\mathcal{U}$. For $\widehat{\mathcal{E}}$, since $\left(I\{\mathcal{P}_{\boldsymbol{\theta}}(\mathbf{x}_i) \geq c\}[\widetilde{m}\{\mathcal{P}_{\boldsymbol{\theta}}(\mathbf{x}_i), \mathcal{P}_{\boldsymbol{\theta}}\} - m\{\mathcal{P}_{\boldsymbol{\theta}}(\mathbf{x}_i), \mathcal{P}_{\boldsymbol{\theta}}\}]\right)_{i=n+1,\dots,N}$ are independent given $\mathcal{L}$ and bounded by $\sup_{\mathbf{x},\boldsymbol{\theta}} |\widetilde{m}\{\mathcal{P}_{\boldsymbol{\theta}}(\mathbf{x}), \mathcal{P}_{\boldsymbol{\theta}}\} - m\{\mathcal{P}_{\boldsymbol{\theta}}(\mathbf{x}), \mathcal{P}_{\boldsymbol{\theta}}\}| = o_p(1)$, we invoke Hoeffding's inequality conditional on $\mathcal{L}$. It follows that $\widehat{\mathcal{E}} = n^{\frac{1}{2}} E_{\mathbf{x}}(I\{\mathcal{P}_{\boldsymbol{\theta}}(\mathbf{x}) \geq c\}[\widetilde{m}\{\mathcal{P}_{\boldsymbol{\theta}}(\mathbf{x}), \mathcal{P}_{\boldsymbol{\theta}}\} - m\{\mathcal{P}_{\boldsymbol{\theta}}(\mathbf{x}), \mathcal{P}_{\boldsymbol{\theta}}\}]) + O_p\{(n/N)^{1/2}\}$. In addition, $\{\widehat{\mathcal{V}}(s, \mathcal{P}_{\boldsymbol{\theta}}) - \mathcal{V}(s, \mathcal{P}_{\boldsymbol{\theta}})\} = O_p(N^{-1/2})$. It follows that

$$\widehat{\mathcal{W}}_\xi(\boldsymbol{\theta}, c) = n^{\frac{1}{2}} \int_c^1 \frac{\widetilde{f}^{(1)}_{\boldsymbol{\theta}}(u)}{\widetilde{f}^{(0)}_{\boldsymbol{\theta}}(u)} du + o_p(1)$$

where $\widetilde{f}^{(a)}_{\boldsymbol{\theta}}(u) = n^{-1} \sum_{i=1}^n K_h\{\mathcal{P}_\theta(\mathbf{x}_i) - u\}\{y_i - m(u, \mathcal{P}_{\boldsymbol{\theta}})\}^a$. Since $\sup_{u,\boldsymbol{\theta}} |\widetilde{f}^{(1)}_{\boldsymbol{\theta}}(u)| + \sup_{u,\boldsymbol{\theta}} |\widetilde{f}^{(0)}_{\boldsymbol{\theta}}(u) - 1| = O_p\{(nh/\log(n))^{-1/2} + n^{\frac{1}{2}} h^2\}$ and $h = o(n^{-1/4})$, we have

$$\widehat{\mathcal{W}}_\xi(\boldsymbol{\theta}, c) = n^{-\frac{1}{2}} \sum_{i=1}^n \int_c^1 K_h\{\mathcal{P}_\theta(\mathbf{x}_i) - u\}\{y_i - m(u, \mathcal{P}_{\boldsymbol{\theta}})\} du + o_p(1) = \mathbb{G}_n\{\mathcal{W}_\xi(\boldsymbol{\theta}, c; \mathbf{D})\} + o_p(1)$$

where $\mathcal{W}_\xi(\boldsymbol{\theta}, c; \mathbf{D}_i) = I\{\mathcal{P}_{\boldsymbol{\theta}}(\mathbf{x}_i) \geq c\}[y_i - m\{\mathcal{P}_{\boldsymbol{\theta}}(\mathbf{x}_i), \mathcal{P}_{\boldsymbol{\theta}})\}]$. It then follows from a functional central limit theorem (Pollard, 1990) that $\widehat{\mathcal{W}}_\xi(\boldsymbol{\theta}, c)$ converges weakly to a zero-mean Gaussian process and possesses stochastic equicontinuity in $(\boldsymbol{\theta}, c)$ under the standard variance metric by Theorem 2.1 of (Kosorok, 2007). Since $\widehat{\boldsymbol{\theta}} \to \boldsymbol{\theta}_0$ in probability, $\{\widehat{\mathcal{W}}_\xi(\widehat{\boldsymbol{\theta}}, c) - \widehat{\mathcal{W}}_\xi(\boldsymbol{\theta}_0, c)\} \to 0$ and hence

$$\widehat{\mathcal{W}}_{\text{TPR}}(c) = n^{\frac{1}{2}} \mu^{-1} \left\{ \widehat{\mathcal{W}}_\xi(\boldsymbol{\theta}_0, c) - \text{TPR}(c) \widehat{\mathcal{W}}_\xi(\boldsymbol{\theta}_0, 0) - \boldsymbol{\psi}_{\mathcal{A}}(\boldsymbol{\theta}_0, c) n^{-1} \sum_{i=1}^n \mathbf{V}_{\boldsymbol{\theta}_{\mathcal{A}}}(\mathbf{D}_i) \right\} + o_p(1)$$
$$= \mathbb{G}_n\{\mathcal{W}_{\text{TPR}}(\boldsymbol{\theta}_0, c; \mathbf{D})\} + o_p(1). \tag{A.1}$$

where $\mathcal{W}_{\text{TPR}}(\boldsymbol{\theta}_0, c; \mathbf{D}) = \mu^{-1}\{\mathcal{W}_\xi(\boldsymbol{\theta}_0, c; \mathbf{D}) - \text{TPR}(c) \mathcal{W}_\xi(\boldsymbol{\theta}_0, 0; \mathbf{D}) - \boldsymbol{\psi}_{\mathcal{A}}(\boldsymbol{\theta}_0, c)^\intercal \mathbf{V}_{\boldsymbol{\theta}_{\mathcal{A}}}(\mathbf{D})\}$, $\boldsymbol{\psi}(\boldsymbol{\theta}, c) = \partial\{\xi(\boldsymbol{\theta}, c) - \text{TPR}(c)\xi(\boldsymbol{\theta}, 0)\}/\partial\boldsymbol{\theta}$ and $\mu = P(y = 1)$.

Analogous arguments can be used to show the uniform consistency of $\widehat{\text{FPR}}(c)$ for $\text{FPR}(c)$

and

$$\widehat{\mathcal{W}}_{\text{FPR}}(c) = n^{1/2}\{\widehat{\text{FPR}}(c) - \text{FPR}(c)\} = \mathbb{G}_n\{\mathcal{W}_{\text{FPR}}(\boldsymbol{\theta}_0, c; \mathbf{D})\} + o_p(1) \tag{A.2}$$

where

$$\mathcal{W}_{\text{FPR}}(\boldsymbol{\theta}_0, c; \mathbf{D}) = -\mu_0^{-1}\left\{\mathcal{W}_\xi(\boldsymbol{\theta}_0, c; \mathbf{D}) - \text{FPR}(c)\mathcal{W}_\xi(\boldsymbol{\theta}_0, 0; \mathbf{D}) - \boldsymbol{\phi}_{\mathcal{A}}(\boldsymbol{\theta}_0, c)^\intercal \mathbf{V}_{\boldsymbol{\theta}_{\mathcal{A}}}(\mathbf{D})\right\}$$

$\boldsymbol{\phi}(\boldsymbol{\theta}, c) = \partial\{\xi(\boldsymbol{\theta}, c) - \text{FPR}(c)\xi(\boldsymbol{\theta}, 0)\}/\partial\boldsymbol{\theta}$, and $\mu_0 = 1 - \mu$. This implies the weak convergence of $\widehat{\mathcal{W}}_{\text{FPR}}(c)$ to a zero-mean Gaussian process. Similar asymptotic properties can be obtained for $\widehat{\text{PPV}}$ and $\widehat{\text{NPV}}$. For $\widehat{\text{ROC}}(u_0) = \widehat{\text{TPR}}\{\widehat{\text{FPR}}^{-1}(u_0)\}$, we note that the uniform consistency of $\widehat{\text{TPR}}(c)$ and $\widehat{\text{FPR}}(c)$ directly implies the uniform consistency of $\widehat{\text{ROC}}(u_0)$ for $\text{ROC}(u_0)$. The weak convergence of $\widehat{\mathcal{W}}_{\text{ROC}}(u_0) = n^{\frac{1}{2}}\{\widehat{\text{ROC}}(u_0) - \text{ROC}(u_0)\}$ also directly follows from the weak convergences of $\widehat{\mathcal{W}}_{\text{TPR}}(c)$ and $\widehat{\mathcal{W}}_{\text{FPR}}(c)$. Specifically from (A.1) and (A.2), we have

$$\begin{aligned}
\widehat{\mathcal{W}}_{\text{ROC}}(u_0) &= \widehat{\mathcal{W}}_{\text{TPR}}(\widehat{c}_{u_0}) + n^{\frac{1}{2}}\left[\text{ROC}\left[\text{FPR}\left\{\widehat{\text{FPR}}^{-1}(u_0)\right\}\right] - \text{ROC}(u_0)\right] \\
&= \widehat{\mathcal{W}}_{\text{TPR}}(c_{u_0}) + n^{\frac{1}{2}}\dot{\text{ROC}}(u_0)\left\{\text{FPR}(c_{u_0}) - \widehat{\text{FPR}}(c_{u_0})\right\} + o_p(1) \\
&= \mathbb{G}_n\left\{\mathcal{W}_{\text{ROC}}(\boldsymbol{\theta}_0, c_{u_0}, \mathbf{D})\right\} + o_p(1). \qquad \text{(A.3)}
\end{aligned}$$

where $\mathcal{W}_{\text{ROC}}(\boldsymbol{\theta}_0, c_{u_0}, \mathbf{D}) = \mathcal{W}_{\text{TPR}}(\boldsymbol{\theta}_0, c_{u_0}, \mathbf{D}) - \dot{\text{ROC}}(u_0)\mathcal{W}_{\text{FPR}}(\boldsymbol{\theta}_0, c_{u_o}, \mathbf{D})$. It follows that $\widehat{\mathcal{W}}_{\text{ROC}}(u_0)$ converges weakly to a zero-mean Gaussian process with variance function $\sigma^2(u_0) = E\{\mathcal{W}_{\text{ROC}}(\boldsymbol{\theta}_0, u_0, \mathbf{D})^2\}$.

Using similar arguments given above, it is straightforward to show that the supervised estimator $\widetilde{\text{ROC}}(u_0)$ is also consistent for $\overline{\text{ROC}}(u_0)$ and $\widetilde{\mathcal{W}}_{\text{ROC}}(u_0) = n^{1/2}\{\widetilde{\text{ROC}}(u_0) - \text{ROC}(u_0)\}$ is asymptotically equivalent to

$$\mathbb{G}_n\left\{\mathcal{W}^{\text{SL}}_{\text{TPR}}(\boldsymbol{\theta}_0, c_{u_0}, \mathbf{D}) - \dot{\text{ROC}}(u_0)\mathcal{W}^{\text{SL}}_{\text{FPR}}(\boldsymbol{\theta}_0, c_{u_o}, \mathbf{D})\right\} \tag{A.4}$$

where

$$\mathcal{W}^{\text{SL}}_{\text{TPR}}(\boldsymbol{\theta}_0, c; \mathbf{D}) = \mu^{-1}\{\mathcal{W}^{\text{SL}}_\xi(\boldsymbol{\theta}_0, c; \mathbf{D}) - \text{TPR}(c)\mathcal{W}^{\text{SL}}_\xi(\boldsymbol{\theta}_0, 0; \mathbf{D}) - \boldsymbol{\psi}_{\mathcal{A}}(\boldsymbol{\theta}_0, c)^\intercal \mathbf{V}_{\boldsymbol{\theta}_{\mathcal{A}}}(\mathbf{D})\},$$

$$\mathcal{W}^{\text{SL}}_{\text{FPR}}(\boldsymbol{\theta}_0, c; \mathbf{D}) = -\mu_0^{-1}\left\{\mathcal{W}^{\text{SL}}_\xi(\boldsymbol{\theta}_0, c; \mathbf{D}) - \text{FPR}(c)\mathcal{W}^{\text{SL}}_\xi(\boldsymbol{\theta}_0, 0; \mathbf{D}) - \boldsymbol{\phi}_{\mathcal{A}}(\boldsymbol{\theta}_0, c)^\intercal \mathbf{V}_{\boldsymbol{\theta}_{\mathcal{A}}}(\mathbf{D})\right\},$$

$$\mathcal{W}^{\text{SL}}_\xi(\boldsymbol{\theta}, c; \mathbf{D}_i) = I\{\mathcal{P}_{\boldsymbol{\theta}}(\mathbf{x}_i) \geq c\}(y_i - \mu).$$

# B. Limiting Distribution of $n^{1/2}\{\widehat{\textbf{ROC}}_{\text{cv}}(u_0) - \textbf{ROC}(u_0)\}$

We now establish that $\widehat{\mathcal{W}}^{\text{cv}}_{\text{ROC}}(u_0) = n^{1/2}\{\widehat{\text{ROC}}_{\text{cv}}(u_0) - \text{ROC}(u_0)\}$ has the same limiting distribution as $\widehat{\mathcal{W}}_{\text{ROC}}(u_0)$. To this end, we derive the asymptotic expansion of $\widehat{\mathcal{W}}^k_{\text{ROC}}(u_0) = n^{1/2}\{\widehat{\text{ROC}}_k(u_0) - \text{ROC}(u_0)\}$ for each partition $\mathcal{I}_k$. Let $\boldsymbol{\zeta} = \{\zeta_i | i = 1, \ldots, n\}$ be $n$ exchangeable discrete random variables uniformly distributed over $\{1, \ldots, \mathcal{K}\}$ independent of the data that satisfy $\sum_{i=1}^n I(\zeta_i = k) = n/\mathcal{K}$, $k = 1, \ldots, \mathcal{K}$. Also let

$$\widehat{\xi}_k(c, \boldsymbol{\theta}) = N^{-1} \sum_{i=1}^N I\{\widehat{\mathcal{P}}_{\boldsymbol{\theta}}(\mathbf{x}_i) \geq c\} \widetilde{m}_k\{\widehat{\mathcal{P}}_{\boldsymbol{\theta}}(\mathbf{x}_i), \widehat{\mathcal{P}}_{\boldsymbol{\theta}}\} = \int_c^1 \widetilde{m}_k(s, \widehat{\mathcal{P}}_{\boldsymbol{\theta}}) \widehat{\mathcal{V}}(ds, \widehat{\mathcal{P}}_{\boldsymbol{\theta}}),$$

and $\widetilde{m}_k(s, \widehat{\mathcal{P}}_{\boldsymbol{\theta}}) = [\sum_{i=1}^n K_h\{\widehat{\mathcal{P}}_{\boldsymbol{\theta}}(\mathbf{x}_i) - s\} I\{\zeta_i = k\} y_i] / [\sum_{i=1}^n K_h\{\widehat{\mathcal{P}}_{\boldsymbol{\theta}}(\mathbf{x}_i) - s\} I\{\zeta_i = k\}]$. It follows from the arguments of Appendix A that conditional on $\boldsymbol{\zeta}$,

$$\widehat{\mathcal{W}}^k_{\xi} = n^{\frac{1}{2}}\{\widehat{\xi}_k(c, \boldsymbol{\theta}) - \xi(c, \boldsymbol{\theta})\} = \mathbb{G}_n\{\mathcal{W}^k_{\xi}(\boldsymbol{\theta}_0, c; \mathbf{D})\}$$

where $\mathcal{W}^k_{\xi}(\boldsymbol{\theta}, c; \mathbf{D}) = \mathcal{K}(I\{\mathcal{P}_{\boldsymbol{\theta}}(\mathbf{x}_i) \geq c, \zeta_i = k\}[y_i - m\{\mathcal{P}_{\boldsymbol{\theta}}(\mathbf{x}_i), \mathcal{P}_{\boldsymbol{\theta}})\}])$. Therefore, $\widehat{\mathcal{W}}^k_{\xi}(\boldsymbol{\theta}, c)$ converges weakly to a zero-mean Gaussian process by the functional central limit theorem (Pollard, 1990). It then follows by Theorem 2.1 of Kosorok (2007) that

$$\widehat{\mathcal{W}}^k_{\text{TPR}}(c) = n^{1/2}\{\widehat{\text{TPR}}_k(c) - \text{TPR}(c)\} = \mathbb{G}_n\{\mathcal{W}^k_{\text{TPR}}(\boldsymbol{\theta}_0, c; \mathbf{D})\} + o_p(1) \tag{B.1}$$

and

$$\widehat{\mathcal{W}}^k_{\text{FPR}}(c) = n^{1/2}\{\widehat{\text{FPR}}_k(c) - \text{FPR}(c)\} = \mathbb{G}_n\{\mathcal{W}^k_{\text{FPR}}(\boldsymbol{\theta}_0, c; \mathbf{D})\} + o_p(1) \tag{B.2}$$

where

$$\mathcal{W}^k_{\text{TPR}}(\boldsymbol{\theta}_0, c; \mathbf{D}) = \mu^{-1}\{\mathcal{W}^k_{\xi}(\boldsymbol{\theta}_0, c; \mathbf{D}) - \text{TPR}(c)\mathcal{W}^k_{\xi}(\boldsymbol{\theta}_0, 0; \mathbf{D}) - \boldsymbol{\psi}_{\mathcal{A}}(\boldsymbol{\theta}_0, c)^{\mathsf{T}} \mathbf{V}^{(\text{-}k)}_{\mathcal{A}}(\mathbf{D})\},$$

$$\mathcal{W}^k_{\text{FPR}}(\boldsymbol{\theta}_0, c; \mathbf{D}) = -\mu_0^{-1}\left\{\mathcal{W}^k_{\xi}(\boldsymbol{\theta}_0, c; \mathbf{D}) - \text{FPR}(c)\mathcal{W}^k_{\xi}(\boldsymbol{\theta}_0, 0; \mathbf{D}) - \boldsymbol{\phi}_{\mathcal{A}}(\boldsymbol{\theta}_0, c)^{\mathsf{T}} \mathbf{V}^{(\text{-}k)}_{\mathcal{A}}(\mathbf{D})\right\},$$

and $\mathbf{V}^{(\text{-}k)}_{\boldsymbol{\theta}_{\mathcal{A}}}(\mathbf{D}_i) = \mathbb{A}_{11}^{-1}\mathbf{U}_{\mathcal{A}}(\boldsymbol{\theta}_0, \mathbf{D}_i)\mathcal{K}I\{\zeta_i \neq k\}/(\mathcal{K}-1)$. Note here that $p$ is the product probability measure generated by $\mathcal{L}$ and $\boldsymbol{\zeta}$. Then from (B.1) and (B.2), we have

$$\begin{aligned}\widehat{\mathcal{W}}^k_{\text{ROC}}(u_0) &= \widehat{\mathcal{W}}^k_{\text{TPR}}(\widehat{c}^{(v)}_{u_0}) + n^{\frac{1}{2}}\left[\text{ROC}\left[\text{FPR}\left\{\widehat{\text{FPR}}_k^{-1}(u_0)\right\}\right] - \text{ROC}(u_0)\right] \\ &= \widehat{\mathcal{W}}^k_{\text{TPR}}(c_{u_0}) + n^{\frac{1}{2}}\dot{\text{ROC}}(u_0)\left\{\text{FPR}(c_{u_0}) - \widehat{\text{FPR}}_k(c_{u_0})\right\} + o_p(1) \\ &= \mathbb{G}_n\left\{\mathcal{W}^k_{\text{ROC}}(\boldsymbol{\theta}_0, c_{u_0}, \mathbf{D})\right\} + o_p(1). \end{aligned} \tag{B.3}$$

where $\widehat{c}^{(v)}_{u_0} = \widehat{\text{FPR}}_k^{-1}(u_0)$ and $\mathcal{W}^k_{\text{ROC}}(\boldsymbol{\theta}_0, c_{u_0}, \mathbf{D}) = \mathcal{W}^k_{\text{TPR}}(\boldsymbol{\theta}_0, c_{u_0}, \mathbf{D}) - \dot{\text{ROC}}(u_0)\mathcal{W}^k_{\text{FPR}}(\boldsymbol{\theta}_0, c_{u_o}, \mathbf{D})$. Since $\sum_{k=1}^{\mathcal{K}} I(\zeta_i = k) = 1$ and $\sum_{k=1}^{\mathcal{K}} I(\zeta_i \neq k) = \mathcal{K}-1$, this implies that

$$\widehat{\mathcal{W}}^{\text{cv}}_{\text{ROC}}(u_0) = \mathcal{K}^{-1}\sum_{k=1}^{\mathcal{K}}\mathbb{G}_n\left\{\mathcal{W}^k_{\text{ROC}}(\boldsymbol{\theta}_0, c_{u_0}, \mathbf{D})\right\} + o_p(1) = \mathbb{G}_n\left\{\mathcal{W}_{\text{ROC}}(\boldsymbol{\theta}_0, c_{u_0}, \mathbf{D})\right\} + o_p(1).$$

Thus $\widehat{\mathcal{W}}^{\text{cv}}_{\text{ROC}}$ is asymptotically equivalent to $\widehat{\mathcal{W}}_{\text{ROC}}$.

# C. Justification for the Resampling Procedure

Here we outline a justification for the proposed resampling procedure. To this end, we consider the unconditional distribution of $\widehat{\mathcal{W}}^*_{\text{ROC}} = n^{1/2}\{\widehat{\text{ROC}}^*(u_0) - \widehat{\text{ROC}}(u_0)\}$. We first note that

$$N^{-1}\sum_{i=n+1}^{N+n} I(\widehat{\mathcal{P}}_{\widehat{\boldsymbol{\theta}}^*i} \geq c)\{\widetilde{m}^*_A(\widehat{\mathcal{P}}_{\widehat{\boldsymbol{\theta}}i}) + \widetilde{m}(\widehat{\mathcal{P}}_{\widehat{\boldsymbol{\theta}}^*i})\} = \hat{\xi}^*(c, \widehat{\boldsymbol{\theta}}) + \hat{\xi}(c, \widehat{\boldsymbol{\theta}}^*) - \hat{\xi}(c, \widehat{\boldsymbol{\theta}}) + o_{p^*}(1)$$

where

$$\begin{aligned}\hat{\xi}^*(c, \boldsymbol{\theta}) &= N^{-1}\sum_{i=1}^{N} I\{\widehat{\mathcal{P}}_{\boldsymbol{\theta}}(\mathbf{x}_i) \geq c\}[\widetilde{m}^*_A\{\widehat{\mathcal{P}}_{\boldsymbol{\theta}}(\mathbf{x}_i), \widehat{\mathcal{P}}_{\boldsymbol{\theta}}\} + \widetilde{m}\{\widehat{\mathcal{P}}_{\boldsymbol{\theta}}(\mathbf{x}_i), \widehat{\mathcal{P}}_{\boldsymbol{\theta}}\}] \\ &= \int_c^1 \widetilde{m}^*_A(s, \widehat{\mathcal{P}}_{\boldsymbol{\theta}})\widehat{\mathcal{V}}(ds, \widehat{\mathcal{P}}_{\boldsymbol{\theta}}) + \widehat{\xi}(c, \boldsymbol{\theta}).\end{aligned}$$

It therefore suffices to derive the asymptotic expansion for $\widehat{\mathcal{W}}^*_{\xi} = n^{1/2}\{\hat{\xi}^*(c, \boldsymbol{\theta}) - \widehat{\xi}(c, \boldsymbol{\theta})\}$. Our previous arguments imply

$$\widehat{\mathcal{W}}^*_{\xi} = \int_c^1 \widetilde{m}^*_A(s, \mathcal{P}_{\boldsymbol{\theta}})\widehat{\mathcal{V}}(ds, \mathcal{P}_{\boldsymbol{\theta}}) = \widehat{\mathcal{E}}^* + n^{\frac{1}{2}}\int_c^1\{\widetilde{m}^*(s, \mathcal{P}_{\boldsymbol{\theta}}) - \widetilde{m}^{\text{cv}}(s, \mathcal{P}_{\boldsymbol{\theta}})\}\widehat{\mathcal{V}}(ds, \mathcal{P}_{\boldsymbol{\theta}}) + o_p(1)$$

where

$$\widehat{\mathcal{E}}^* = n^{\frac{1}{2}}\mathbb{P}_N[I\{\mathcal{P}_{\boldsymbol\theta}(\mathbf{x}) \geq c\}\{\widetilde{m}^{\text{cv}}(s,\mathcal{P}_{\boldsymbol\theta}) - \widetilde{m}^*_k(s,\mathcal{P}_{\boldsymbol\theta})\}], \quad \widetilde{m}^{\text{cv}}(s,\mathcal{P}_{\boldsymbol\theta}) = \mathcal{K}^{-1}\sum_{k=1}^{\mathcal{K}} \widetilde{m}_{(\text{-}k)}(s,\mathcal{P}_{\boldsymbol\theta}),$$

$$\widetilde{m}^*_k(s,\mathcal{P}_{\boldsymbol\theta}) = \left(\sum_{k=1}^{\mathcal{K}}\sum_{i\in\mathbb{I}_k} K_h\{\mathcal{P}_{\boldsymbol\theta}(\mathbf{x}_i) - s\}[\widetilde{m}_{(\text{-}k)}\{\mathcal{P}_{\boldsymbol\theta}(\mathbf{x}_i)\}G_i]\right) \Big/ \left[\sum_{i=1}^n K_h\{\mathcal{P}_{\boldsymbol\theta}(\mathbf{x}_i) - s\}G_i\}\right],$$

$$\text{and} \quad \widetilde{m}^*(s,\mathcal{P}_{\boldsymbol\theta}) = \left[\sum_{i=1}^n K_h\{\mathcal{P}_{\boldsymbol\theta}(\mathbf{x}_i) - s\}G_i y_i\right] \Big/ \left[\sum_{i=1}^n K_h\{\mathcal{P}_{\boldsymbol\theta}(\mathbf{x}_i) - s\}G_i\}\right].$$

For $\widehat{\mathcal{E}}^*$, since $n^{-1}\sum_{i=1}^n K_h\{\mathcal{P}_{\boldsymbol\theta}(\mathbf{x}_i) - s\}G_i \geq \gamma > 0$,

$$\widehat{\mathcal{E}}^* \leq n^{1/2}\gamma^{-1}\int_c^1 \tilde{f}^*_{\boldsymbol\theta}(u)du$$

where $\tilde{f}^*_{\boldsymbol\theta}(u) = n^{-1}\sum_{k=1}^{\mathcal{K}}\sum_{i\in\mathcal{I}_k} K_h\{\mathcal{P}_{\boldsymbol\theta}(\mathbf{x}_i) - u\}G_i[\widetilde{m}_{(\text{-}k)}\{\mathcal{P}_{\boldsymbol\theta}(\mathbf{x}_i)\} - \widetilde{m}_{(\text{-}k)}(u,\mathcal{P}_{\boldsymbol\theta})]$. It follows that $\widehat{\mathcal{E}}^* = o_{p^*}(1)$ as $\sup_{u,\boldsymbol\theta}|\tilde{f}^*_{\boldsymbol\theta}(u)| = O_{p^*}(h^2)$ and $h = o(n^{-1/4})$. On the other hand, the arguments in Appendices A and B can be applied to show that

$$\widehat{\mathcal{W}}^*_\xi = \int_c^1 \{\widetilde{m}^*(s,\mathcal{P}_{\boldsymbol\theta}) - \widetilde{m}^{\text{cv}}(s,\mathcal{P}_{\boldsymbol\theta})\}\widehat{\mathcal{V}}(ds,\mathcal{P}_{\boldsymbol\theta}) + o_{p^*}(1) = \mathbb{G}_n\{\mathcal{W}^*_\xi(\boldsymbol\theta, c; \mathbf{D}, G)\} + o_{p^*}(1)$$

where $\mathcal{W}^*_\xi(\boldsymbol\theta, c; \mathbf{D}_i, G_i) = I\{\mathcal{P}_{\boldsymbol\theta}(\mathbf{x}_i) \geq c\}[y_i - m\{\mathcal{P}_{\boldsymbol\theta}(\mathbf{x}_i), \mathcal{P}_{\boldsymbol\theta}\}](G_i - 1)$. Additionally, our arguments in Appendix A verify that $n^{\frac{1}{2}}\{\hat{\xi}(c,\boldsymbol\theta) - \xi(c,\boldsymbol\theta)\}$ converges weakly to a zero-mean Guassian process in $c$ and $\boldsymbol\theta$ and hence $n^{\frac{1}{2}}\{\hat{\xi}(c,\widehat{\boldsymbol\theta}^*) - \hat{\xi}(c,\widehat{\boldsymbol\theta})\} = n^{\frac{1}{2}}\{\xi(c,\widehat{\boldsymbol\theta}^*) - \xi(c,\widehat{\boldsymbol\theta})\} + o_{p^*}(1)$.

By the functional central limit theorem (Pollard, 1990), $\widehat{\mathcal{W}}^*_\xi(\boldsymbol\theta, c)$ converges weakly to a zero-mean Gaussian process. It follows by Theorem 2.1 of Kosorok (2007)

$$\begin{aligned}\widehat{\mathcal{W}}^*_{\text{TPR}}(c) &= n^{1/2}\{\widehat{\text{TPR}}^*(c) - \widehat{\text{TPR}}(c)\} = \mathbb{G}_n\{\mathcal{W}^*_{\text{TPR}}(\boldsymbol\theta_0, c; \mathbf{D}, G)\} + o_{p^*}(1) \quad \text{and} \\ \widehat{\mathcal{W}}^*_{\text{FPR}}(c) &= n^{1/2}\{\widehat{\text{FPR}}^*(c) - \widehat{\text{FPR}}(c)\} = \mathbb{G}_n\{\mathcal{W}^*_{\text{FPR}}(\boldsymbol\theta_0, c; \mathbf{D}, G)\} + o_{p^*}(1) \qquad \text{(C.1)}\end{aligned}$$

where $\mathcal{W}^*_{\text{TPR}}(\boldsymbol\theta_0, c; \mathbf{D}, G) = \{\mathcal{W}^*_\xi(\boldsymbol\theta_0, c; \mathbf{D}, G) - \text{TPR}(c)\mathcal{W}^*_\xi(\boldsymbol\theta_0, 0; \mathbf{D}, G) - \boldsymbol\psi_{\mathcal{A}}(\boldsymbol\theta_0, c)^{\mathsf{T}}\mathbf{V}_{\mathcal{A}}(\mathbf{D})(G - 1)\}/\mu$ and $\mathcal{W}^*_{\text{FPR}}(\boldsymbol\theta_0, c; \mathbf{D}, G) = -\{\mathcal{W}^*_\xi(\boldsymbol\theta_0, c; \mathbf{D}, G) - \text{FPR}(c)\mathcal{W}^*_\xi(\boldsymbol\theta_0, 0; \mathbf{D}, G) - \boldsymbol\phi_{\mathcal{A}}(\boldsymbol\theta_0, c)^{\mathsf{T}}\mathbf{V}_{\mathcal{A}}(\mathbf{D})$ $(G - 1)\}/\mu_0$. Then from (C.1), we have

$$\begin{aligned}\widehat{\mathcal{W}}^*_{\text{ROC}}(u_0) &= \widehat{\mathcal{W}}^*_{\text{TPR}}(\widehat{c}^*_{u_0}) + n^{\frac{1}{2}}\left(\text{ROC}\left[\text{FPR}\left\{(\widehat{\text{FPR}}^*)^{-1}(u_0)\right\}\right] - \text{ROC}(u_0)\right) \\ &= \widehat{\mathcal{W}}^*_{\text{TPR}}(c_{u_0}) + \dot{\text{ROC}}(u_0)\left\{\text{FPR}(c_{u_0}) - \widehat{\text{FPR}}^*(c_{u_0})\right\} + o_p(1)\end{aligned}$$

$$= \mathbb{G}_n \left\{ \mathcal{W}^*_{\mathrm{TPR}}(\boldsymbol{\theta}_0, c_{u_0}, \mathbf{D}, G) - \dot{\mathrm{ROC}}(u_0) \mathcal{W}^*_{\mathrm{FPR}}(\boldsymbol{\theta}_0, c_{u_o}, \mathbf{D}, G) \right\} + o_{p^*}(1). \quad \text{(C.2)}$$

where $\widehat{c}^*_{u_0} = \widehat{\mathrm{FPR}}^{*^{-1}}(u_0)$.By the multiplier central limit theorem (Van Der Vaart and Wellner, 1996), the distribution of $\widehat{\mathcal{W}}^*_{\mathrm{ROC}}(u_0)|\mathcal{L}$ converges to a zero-mean normal random variable. This then implies that for $\epsilon > 0$ there exists $\mathcal{M}$ such that for $n > \mathcal{M}$ the probability of

$$\sup_{v \in \mathbb{R}} |P(\widehat{\mathcal{W}}^*_{\mathrm{ROC}}(u_0) \le v | \mathcal{L}) - P(\widehat{\mathcal{W}}_{\mathrm{ROC}}(u_0) \le v)| < \epsilon$$

with respect to $\mathcal{L}$ is at least $1 - \epsilon$. This justifies the proposed resampling procedure.

# REFERENCES

Kosorok, M. R. (2007) *Introduction to empirical processes and semiparametric inference.* Springer Science & Business Media.

Pollard, D. (1990) Empirical processes: Theory and applications. IMS.

Van Der Vaart, A. W. and Wellner, J. A. (1996) *Weak Convergence.* Springer.